%% file: main.tex
\documentclass[lettersize,journal]{IEEEtran}

\usepackage{amsmath,amsfonts}
\usepackage{algorithmic}
\usepackage{array}
\usepackage[caption=false,font=normalsize,labelfont=sf,textfont=sf]{subfig}
\usepackage{textcomp}
\usepackage{stfloats}
\usepackage{url}
\usepackage{verbatim}
\usepackage{graphicx}

\usepackage{capt-of}
\usepackage{multirow}
\usepackage{booktabs}
\usepackage{adjustbox}
\usepackage{float}

\usepackage{colortbl}
\usepackage{wrapfig}
\usepackage[dvipsnames]{xcolor}
\usepackage[normalem]{ulem}
\usepackage{times}
\usepackage{hyperref}
\usepackage{xspace}
\usepackage{chngcntr}
\usepackage[most]{tcolorbox}
\usepackage{cuted}
\usepackage{placeins}

\usepackage[numbers,sort&compress]{natbib}

\definecolor{darkgreen}{rgb}{0.0, 0.5, 0.0} 
\def\benchmark{T2A-bench}
\def\dataset{IF-caps-Pro}

\newcommand{\model}{AudioX-Turbo\xspace}
\newcommand{\modelbase}{AudioX-Base\xspace}

\newcommand{\eg}{\emph{e.g.}} 

\newcommand{\ie}{\emph{i.e.}}

\def\BibTeX{{\rm B\kern-.05em{\sc i\kern-.025em b}\kern-.08em
    T\kern-.1667em\lower.7ex\hbox{E}\kern-.125emX}}
\usepackage{balance}

\begin{document}

\title{AudioX-Turbo: A Unified Framework for Efficient Anything-to-Audio Generation}

\author{
Zeyue~Tian$^{*1,3}$, Lei~Ke$^{*1,2}$, Zhaoyang~Liu$^{1}$, Ruibin~Yuan$^{1}$, Liumeng~Xue$^{1}$, Yujiu~Yang$^{\dagger 2}$,\\
Weijia~Chen$^{3}$, Xu~Tan$^{4}$, Qifeng~Chen$^{1}$, Wei~Xue$^{\dagger 1}$, and~Yike~Guo$^{\dagger 1}$\\[3pt]
{\normalfont\normalsize
$^{1}$The Hong Kong University of Science and Technology\quad
$^{2}$Tsinghua University\\
$^{3}$Noiz AI\quad
$^{4}$Independent Researcher\\[2pt]
}
{\footnotesize $^{*}$Equal contribution.\quad $^{\dagger}$Corresponding authors.}
\thanks{Manuscript received \today; revised \today.}
\thanks{$^{*}$Equal contribution. $^{\dagger}$Corresponding authors: Wei Xue (weixue@ust.hk), Yike Guo (yikeguo@ust.hk).}
\thanks{Z. Tian, Z. Liu, R. Yuan, L. Xue, Q. Chen, W. Xue, and Y. Guo are with the Hong Kong University of Science and Technology, Hong Kong SAR, China. L. Ke and Y. Yang are with Tsinghua University, China. W. Chen is with Noiz AI. X. Tan is an independent researcher.}
}

\markboth{}
{Tian \MakeLowercase{\textit{et al.}}: AudioX-Turbo: A Unified Framework for Efficient Anything-to-Audio Generation}

\twocolumn[{
\begin{@twocolumnfalse}
\maketitle

\vspace{-0.6\baselineskip}
\centering
\includegraphics[width=\textwidth]{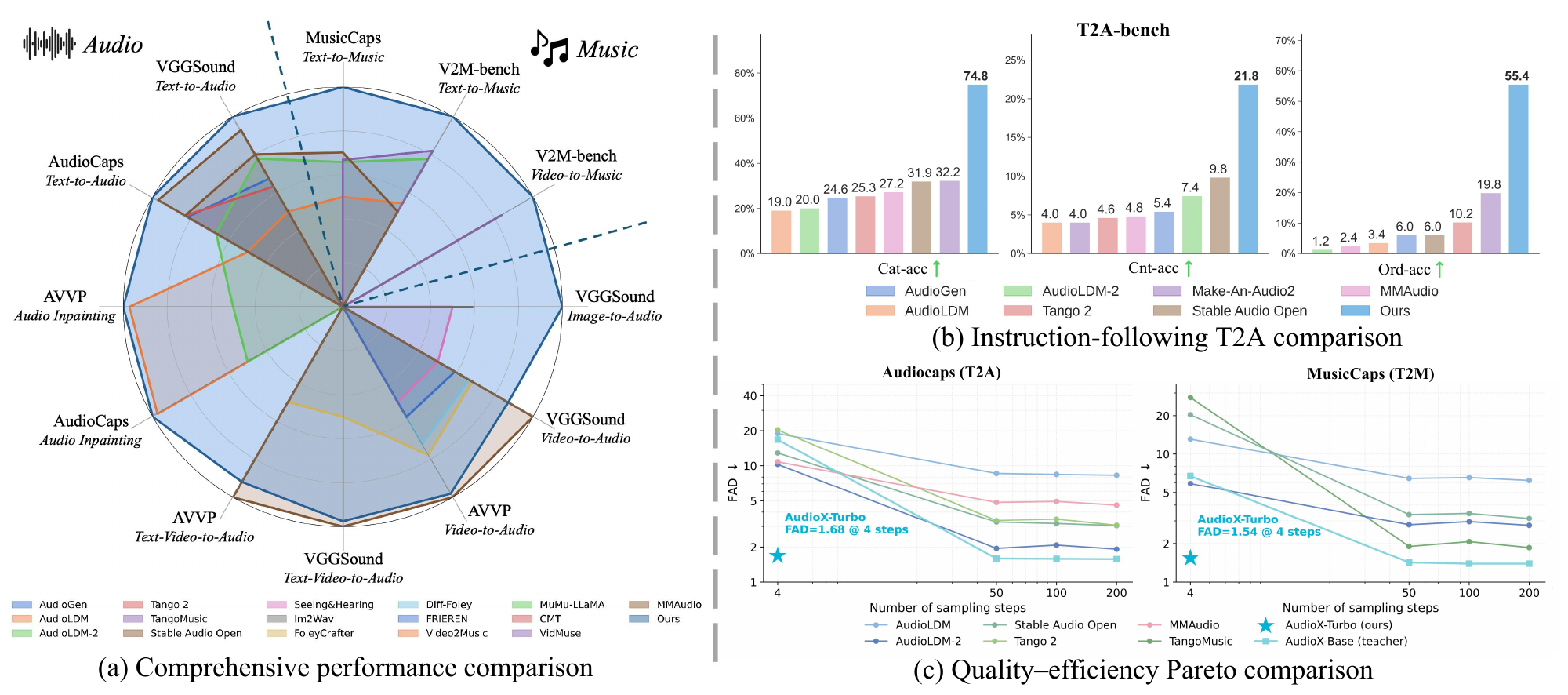}

\captionof{figure}{\textbf{Performance comparison of \model\ against baselines.}
(a) Comprehensive comparison across multiple benchmarks via Inception Score.
(b) Results on instruction-following benchmark.
(c) Quality--efficiency trade-off across diffusion-based methods.
}
\label{fig:teaser}
\vspace{-0.1\baselineskip}

\end{@twocolumnfalse}
}]

\input{sec/0_abstract}

\begin{IEEEkeywords}
Audio Generation, Diffusion Model, Efficient Inference.
\end{IEEEkeywords}

\input{sec/1_Introduction}

\input{sec/2_Related_Work}

\input{sec/3_Dataset}

\input{sec/4_Pretrain}

\input{sec/5_Step_Distillation}

\input{sec/6_Experiment}

\input{sec/7_Conclusion}

\bibliographystyle{IEEEtran}
\bibliography{main}

\input{sec/appendix}

\end{document}

%% file: sec/0_abstract.tex
\begin{abstract}

    Audio and music generation based on flexible multimodal control signals is a widely applicable topic, with the following key challenges: 1) a unified multimodal modeling framework, 2) large-scale, high-quality training data, and 
    3) the prohibitive inference cost of multi-step diffusion sampling.
    As such, we propose \model, a unified and efficient framework for anything-to-audio generation that integrates varied multimodal conditions (\ie, text, video, and audio signals) in this work.
    \model\ follows a \emph{teacher--student} paradigm. The teacher \modelbase\ is built on a Multimodal Diffusion Transformer with a Multimodal Adaptive Fusion module that aligns diverse multimodal inputs for high-fidelity synthesis, and is then distilled into the few-step student \model\ via Distribution Matching Distillation adapted to flow matching, complemented by a diffusion-based discriminator for high-quality few-step generation.
    To support the training of \model, we construct a large-scale, high-quality dataset, \dataset, comprising approximately 9.2M samples curated through a two-stage data collection and annotation pipeline. We benchmark \model\ across a wide range of tasks, finding that our model achieves superior performance, especially on text-to-audio and text-to-music generation, while operating at only \textbf{4 sampling steps} and requiring up to $\sim$25$\times$ fewer function evaluations (NFE) than multi-step baselines. These results demonstrate that our method is capable of audio generation under flexible multimodal control, showing efficient and powerful instruction-following capabilities.
    The code and datasets will be available at \url{https://zeyuet.github.io/AudioX-Turbo/}.

\end{abstract}

%% file: sec/1_Introduction.tex
\section{Introduction}
\label{sec:intro}
\IEEEPARstart{I}{n} recent years, audio generation, especially for sound effects and music, has emerged as a crucial component in multimedia creation, showing practical values in enhancing user experiences across a wide range of applications. For example, in social media, film production, and video games, sound effects and music significantly intensify emotional resonance and engagement with the audience. The ability to create high-quality audio not only enriches multimedia content but also opens up new avenues for creative expression.

However, the manual audio production is time-consuming and requires specialized skills, presenting a compelling research opportunity to automate audio generation. Despite notable advancements~\citep{liu2023audioldm, copet2024simple, wang2024frieren}, the field has predominantly focused on specialized models with constrained inputs and outputs. These models often operate with a single conditioning modality, such as text-to-audio or video-to-audio, and are typically restricted to a single output domain, like generating either sound effects~\citep{cheng2025mmaudio} or music~\citep{tian2025vidmuse} exclusively. While a recent trend towards unification is emerging, with some pioneering works accommodating multiple inputs~\citep{polyak2024movie, zhang2024foleycrafter}, they often lack the flexibility to support diverse modal combinations and exhibit weak instruction-following abilities. As a result, the potential of unified models still remains underexplored. We find that a major factor behind these limitations is the scarcity of high-quality, multimodal data suitable for training unified systems. Existing datasets are often task-specific, typically providing supervision for only one conditioning modality, such as text-to-audio~\citep{kim2019audiocaps}, video-to-audio~\citep{chen2020vggsound}, or video-to-music~\citep{tian2025vidmuse}. This lack of datasets with diverse and combinable control signals has significantly hindered the training of unified models.

Beyond the architectural and data bottlenecks, another often overlooked obstacle is \emph{inference efficiency}. State-of-the-art audio generation models~\citep{liu2023audioldm, evans2024stable, cheng2025mmaudio} typically rely on diffusion or flow matching, requiring tens to over a hundred sequential function evaluations to solve the underlying ODE. Such a high sampling cost leads to substantial inference latency, limiting their applicability to real-time scenarios such as interactive content creation and on-the-fly video-to-audio generation.
While step-distillation techniques~\citep{pd, lcm, dmd, dmd2} have substantially accelerated visual generation, their application to multimodal-conditioned audio generation remains underexplored. In this setting, aggressive few-step sampling can tend to undermine cross-modal alignment and instruction following, which are both critical for controllable audio generation.

To this end, we propose \model, a unified framework for efficient  anything-to-audio generation. We first pretrain a multi-step multimodal teacher, \modelbase, for high-fidelity audio synthesis, and then distill it into an efficient few-step student.
Both \modelbase\ and \model\ share a Transformer-based backbone. Specifically, we adopt a Multimodal Diffusion Transformer (MMDiT) architecture, which unifies multimodal conditioning signals~\citep{wu2023next, liu2024visual, lin2023video} while retaining the high-fidelity generative capability for audio synthesis~\citep{evans2024stable, evans2024long, majumder2024tango}.
To further enhance multimodal representation alignment, we introduce a lightweight Multimodal Adaptive Fusion module that adaptively weights and aligns conditioning modalities before fusion, enabling stronger cross-modal control with significant improvements in audio quality.

To enable efficient few-step inference, we adopt a Distribution Matching Distillation framework~\citep{dmd, dmd2} adapted to the flow matching formulation, complemented by a diffusion-based discriminator that reuses the teacher's multimodal features to preserve cross-modal alignment under aggressive few-step regimes. As a result, \model\ achieves generation quality comparable to the multi-step teacher while enabling substantially faster inference.

To overcome data scarcity, we develop a \emph{two-stage} data construction pipeline. \emph{Stage~1} curates large-scale source pairs: a carefully designed pipeline yields V2M-500K for video-music, complemented by VGGSound~\citep{chen2020vggsound} and AudioSet-Strong~\citep{hershey2021benefit} as video-audio sources. \emph{Stage~2} produces fine-grained multimodal supervision through a Gemini~2.5~Pro plus Qwen2-Audio annotation cascade. The resulting dataset, \dataset, contains approximately 1.3M general audio samples and 7.9M music samples, providing the necessary training signals for unified anything-to-audio modeling.

Trained on our large-scale dataset with the unified design, our model demonstrates exceptional performance and strong instruction-following ability. To validate its effectiveness, we benchmark it against state-of-the-art methods across a comprehensive suite of tasks and established benchmarks. In addition, to rigorously evaluate its instruction-following ability on T2A tasks, we construct a new benchmark, \benchmark. As demonstrated in Sec.~\ref{sec:main_results}, \model\ achieves state-of-the-art or comparable results across multiple benchmarks and tasks while substantially outperforming prior methods in instruction-following capabilities. 
Notably, using only 4 sampling steps, \model\ achieves performance comparable to the multi-step teacher \modelbase, while reducing the number of function evaluations (NFE) by up to approximately $25\times$.
A notable finding from our unified training approach is that we observe a \textbf{\emph{cross-modal regularization effect}} under unified training: improving the quality and granularity of textual supervision leads to better modality alignment, which jointly boosts performance across conditioning modalities (see Sec.~\ref{sec:ablation_study}). This observation provides empirical insight for future multimodal audio generation.

In summary, the main contributions of this work are as follows:

1) We propose \model, a unified and efficient framework for anything-to-audio generation. It supports both audio and music generation from diverse multimodal conditions, relaxing the input-output constraints of task-specific systems. For efficient inference, we distill a multi-step teacher \modelbase\ into a few-step student via Distribution Matching Distillation adapted to flow matching, offering a practical recipe for efficient generalist audio generation.

2) To overcome data scarcity for unified training, we design a \emph{two-stage} data curation and annotation pipeline that aggregates video-audio and video-music sources and produces fine-grained multimodal supervision at scale. This yields \dataset, a large-scale, high-quality dataset of approximately 9.2M samples in total, providing a unified foundation for multimodal-conditioned audio generation.

3) We conduct comprehensive experiments on a wide array of tasks, systematically benchmarking state-of-the-art methods categorized by their input modalities and output domains. Results demonstrate \model's strong multi-task capability and superior instruction-following ability, while matching its multi-step teacher \modelbase\ with up to $\sim$25$\times$ fewer NFE using only 4 sampling steps.

%% file: sec/2_Related_Work.tex
\section{Related work}

\textbf{Audio and music generation.}
Deep generative models~\cite{evans2026stableaudio3,esser2024scaling,huang2023make,liu2023interngpt,liu2024controlllm,liu2025scalecua,tian2024visual,cheng2025mmaudio} have greatly advanced the development of audio and music synthesis. However, most existing methods remain confined to a single modality or support only limited types of conditions. For instance, \emph{text-to-audio} approaches~\citep{liu2023audioldm,majumder2024tango,evans2024long,evans2024stable,jiang2025freeaudio,huang2023make,hung2024tangoflux,mei2026dasheng,tian2026audio,he2024llms} focus on generating diverse soundscapes from textual prompts, while \emph{text-to-music} systems~\citep{copet2024simple,ghosal2023text,yuan2024chatmusician,yuan2025yue,ma2024foundation,deng2024composerx} specialize in composing coherent musical pieces. Separate lines of work tackle tasks like \emph{audio inpainting}~\citep{liu2023audioldm,liu2024audioldm}, primarily with text conditioning.
Meanwhile, \emph{video-to-audio} methods~\citep{zhang2024foleycrafter,luo2024diff,wang2024frieren,polyak2024movie,chen2024video,dai2026omni2sound,hadjeres2026woosh} typically generate foley or environmental sounds synchronized to visual cues. Some of these also incorporate text for additional context, thereby bridging visual and textual modalities. Beyond sound effects, \emph{video-to-music} approaches~\citep{kang2024video2music,liu2024mumu,di2021video,tian2025vidmuse,li2024diff,lin2024vmas,li2024muvi,liu2025unimoe} align musical compositions with the visual content to enhance narrative depth in multimedia applications.
Despite these advances, current frameworks often specialize in only one modality or rely on a limited set of input conditions, hindering multi-task adaptation and restricting their ability to scale or transfer knowledge across related tasks. In contrast, our \emph{unified} approach supports both audio and music generation for a broad range of input conditions—including text, video, and audio—all within a single framework.

\noindent\textbf{Audio Datasets.}
While substantial research efforts have led to the creation of valuable datasets for specific tasks like text-to-audio \citep{kim2019audiocaps, xue2025audio, drossos2020clotho, wu2023large}, text-to-music \citep{copet2024simple,liu2024mumu, ramires2020freesound}, video-to-audio \citep{chen2020vggsound,hershey2021benefit,tian2020unified}, and video-to-music \citep{tian2025vidmuse, zhou2025harmonyset, chi2024mmtrail, chi2026vmchill}, training a generalist unified model remains under explored. The existing training data is typically constrained to a single conditioning modality and a narrow output domain (e.g., only sound effects or only music). It has significantly hindered progress towards developing more versatile and robust systems. To overcome this critical data scarcity, we introduce a large-scale, multimodal dataset constructed via a novel annotation and augmentation pipeline, specifically designed to provide the comprehensive supervision required for unified audio and music generation.

\noindent\textbf{Diffusion models.}
Denoising diffusion models \citep{ho2020denoising,song2020score} have become a cornerstone of modern generative modeling, achieving state-of-the-art results in image \citep{rombach2022high,ramesh2022hierarchical,brooks2023instructpix2pix}, video \citep{chen2023videocrafter1,ho2022video,guo2023animatediff}, and audio synthesis \citep{popov2021grad,jeong2021diff,liu2024audioldm,liu2023audioldm,liu2022diffsinger,evans2024long,cheng2025mmaudio}. However, their application in the audio domain has predominantly been limited to single-condition tasks (e.g., text-to-audio), falling short of the more generalized ``anything-to-audio'' scenarios where inputs can be multimodal. To bridge this gap, our framework leverages diffusion models for multi-condition generation, offering a more flexible and universal paradigm.

\begin{figure*}[t]
    \centering
    \includegraphics[width=\linewidth]{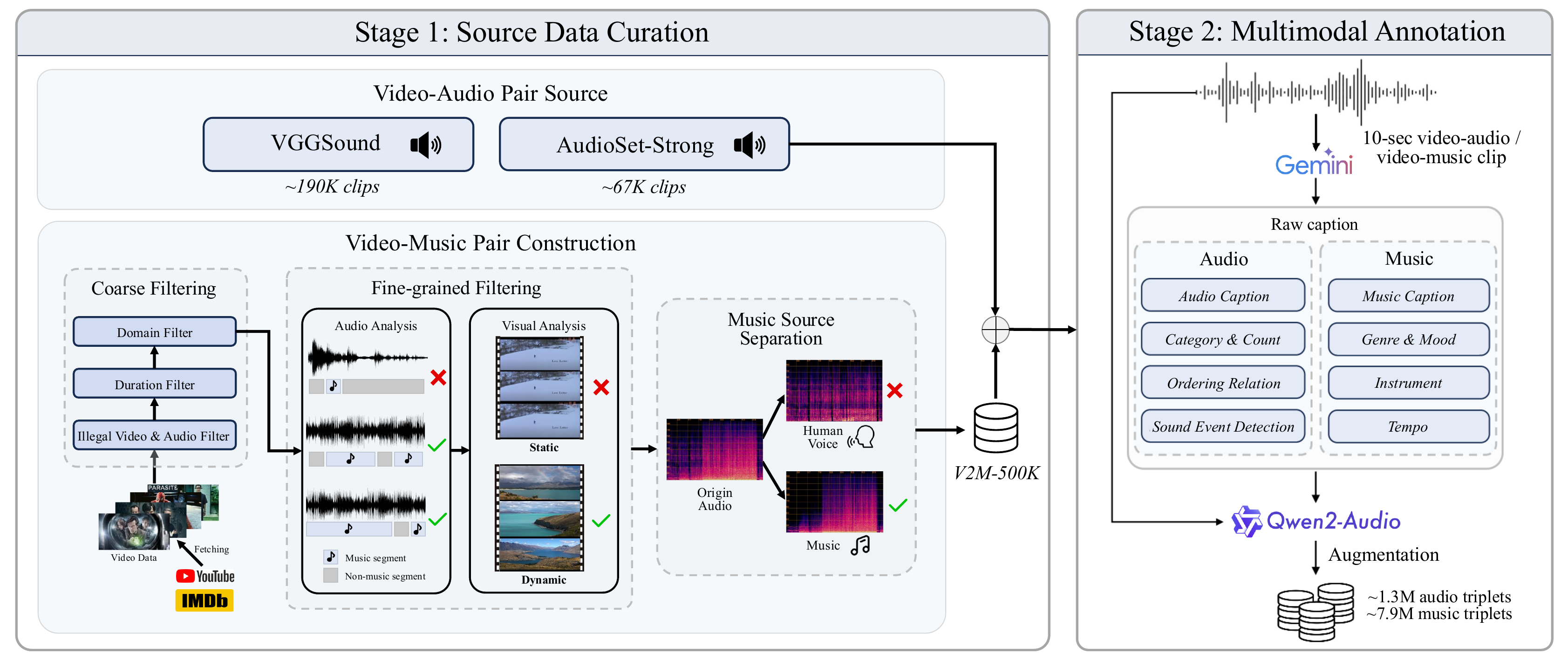}
    \caption{\textbf{Two-stage data construction pipeline of \dataset.} 
    \emph{Stage~1} curates video-audio (VGGSound, AudioSet-Strong) and video-music (V2M-500K) source pairs. 
    \emph{Stage~2} enriches them with fine-grained annotations via a Gemini~2.5~Pro and Qwen2-Audio annotation cascade, producing $\sim$1.3M video-text-audio and $\sim$8M video-text-music triplets.}
    \label{fig:dataset_pipeline}
\end{figure*}

\noindent\textbf{Diffusion Acceleration.}
Step distillation has emerged as a primary strategy for reducing the high sampling cost of diffusion models. \emph{Trajectory-preserving} methods aim to match the teacher's generation path with fewer steps, including Progressive Distillation~\cite{pd}, Consistency Distillation and its trajectory-aware variants~\cite{cm, lcm, ctm, pcm}, and Rectified Flow~\cite{flowmatch, flowsteer, proreflow} that straightens ODE trajectories. More recent \emph{distribution-matching} methods relax this constraint and directly align the student's output distribution with the target, via score-based objectives such as Distribution Matching Distillation (DMD)~\cite{dmd, dmd2}, achieving stronger few-step quality. While these techniques are well-developed for image generation, their extension to multimodal audio generation remains largely unexplored. We adopt a distribution-matching formulation tailored to flow matching to accelerate our pre-trained audio generation model.

%% file: sec/3_Dataset.tex
\section{Dataset}
\label{sec:dataset}

Training a unified anything-to-audio model is bottlenecked by two complementary data gaps. 
First, while \emph{video-audio} pairs are well covered by curated public corpora~\citep{chen2020vggsound,hershey2021benefit}, large-scale and high-quality \emph{video-music} datasets remain scarce, with most existing resources suffering from limited scale, narrow genre coverage, or data quality issues~\citep{hong2017content,zhuo2023video,tian2025vidmuse}. 
Second, even when raw paired data is available, existing audio datasets generally lack the high-quality, multimodal conditioning signals (e.g., fine-grained captions) necessary to train versatile, unified models. To bridge both gaps, we construct \dataset\ through a \emph{two-stage} pipeline (Fig.~\ref{fig:dataset_pipeline}): \emph{Stage~1} curates large-scale source video-audio and video-music pairs, and \emph{Stage~2} produces fine-grained multimodal annotations on top of them.

\begin{figure}[t]
    \centering
    \includegraphics[width=\linewidth]{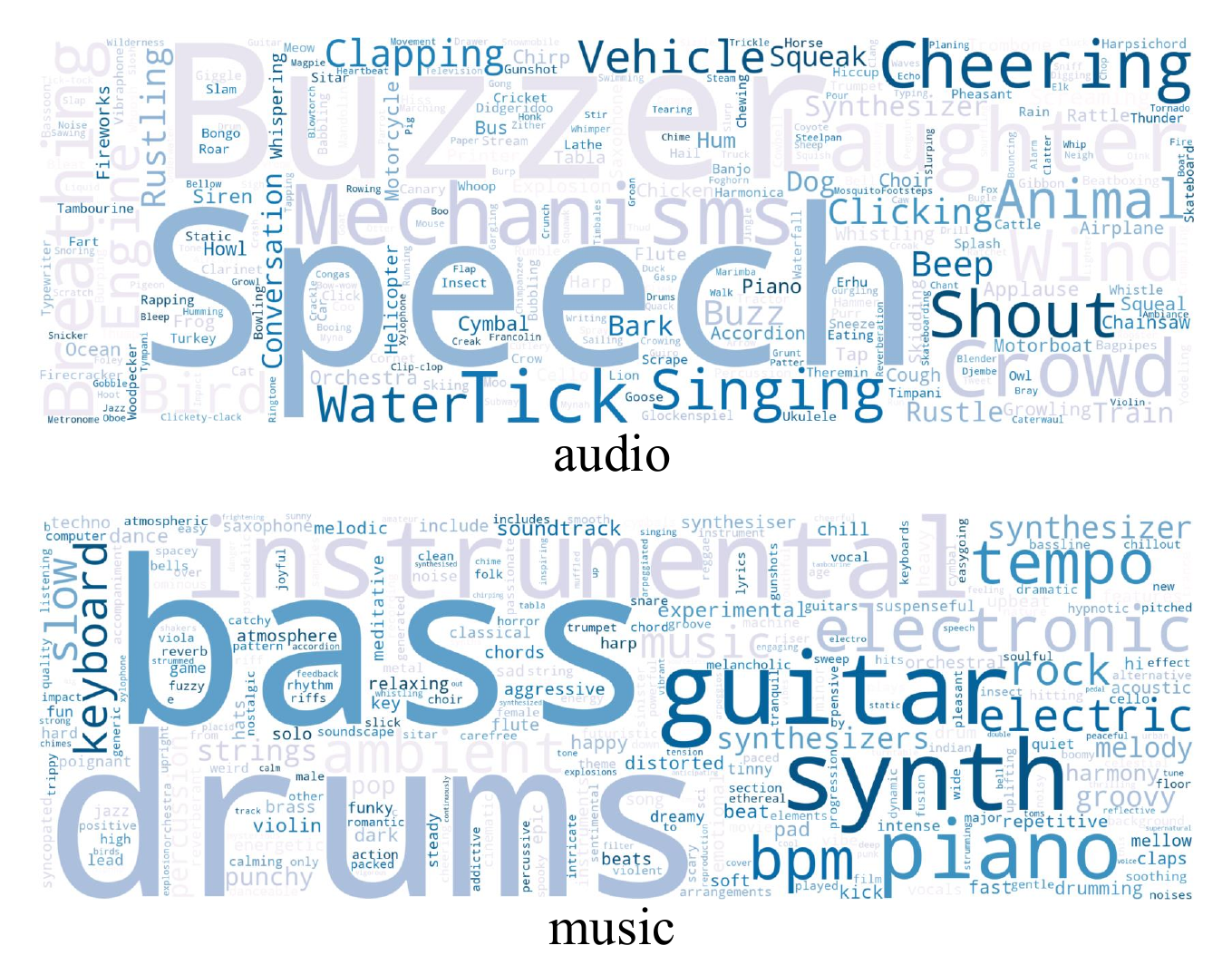}
    \caption{\textbf{Word clouds of \dataset.} Most frequent terms in our curated captions for the general-audio (top) and music (bottom) domains, illustrating the diversity of the annotations.}
    \label{fig:dataset}
\end{figure}

\subsection{Stage 1: Source Data Curation}
\label{sec:source_data}

For \emph{video-audio} pairs, we directly leverage VGGSound~\citep{chen2020vggsound} and AudioSet-Strong~\citep{hershey2021benefit}, two large-scale public corpora that have undergone rigorous curation and provide reliable event-level category labels, which serve as grounding keywords for the LLM-based annotation in Stage~2.
For \emph{video-music} pairs, we construct \textbf{V2M-500K}, a large-scale corpus of high-quality video-music pairs, through a multi-step collection-and-filtering pipeline. We first design a set of YouTube and IMDb-derived queries to retrieve videos whose visuals are tightly coupled with music, spanning a wide range of video types such as movie trailers, advertisements, documentaries, and vlogs. As the raw collection inevitably contains noisy samples, we apply a cascade of filters to obtain reliable video-music pairs. \emph{Coarse filtering} removes videos with broken audio or video tracks, unsuitable duration, inappropriate content and noising background music uncorrelated to the visuals (\eg, interviews, news). \emph{Fine-grained filtering} then keeps videos with substantial musical content and dynamic visuals: a pretrained audio classifier~\citep{kong2020panns} detects music segments, and a perceptual quality model discards visually static or low-quality clips. Finally, we use \emph{music source separation} to isolate the music track from speech and ambient sounds, yielding clean video-music pairs. Detailed statistics, genre distribution, and additional construction protocols are provided in Appendix~\ref{sec:appendix_datasets}.

\subsection{Stage 2: Multimodal Annotation Pipeline}
\label{sec:if_caps_xl}

The source pairs from Stage~1 still lack the rich, detailed textual supervision required for training versatile, unified models. We therefore design a two-step LLM-based annotation pipeline that produces a global caption together with task-specific structured fields for every 10-second clip. \textbf{\emph{First}}, we employ a powerful multimodal LLM (Gemini~2.5 Pro) to generate a comprehensive set of initial annotations for each 10-second clip. These annotations consist of a holistic global caption and a set of structured fields: for general audio, these fields include sound event classification and count; for music, they specify attributes like genre and instrumentation. \textbf{\emph{Then}}, since using the resource-intensive Gemini model for the entire dataset is costly, we leverage the open-source Qwen2-Audio~\citep{chu2024qwen2} model to augment these structured fields at a large scale. Conditioned on both the initial annotations and the raw audio, the model generates varied captions, enhancing data diversity while managing costs. This process yields comprehensive, fine-grained captions for approximately 1.3M video-text-audio triplets and 7.9M video-text-music triplets. The diversity of our curated dataset is highlighted by the word clouds in Fig.~\ref{fig:dataset}. More details and samples of our annotated data are provided in the Appendix~\ref{sec:appendix_if-caps}.

%% file: sec/4_Pretrain.tex
\section{Unified Anything-to-Audio Pretraining}

\begin{figure*}[t]
    \centering
    \includegraphics[width=1\textwidth]{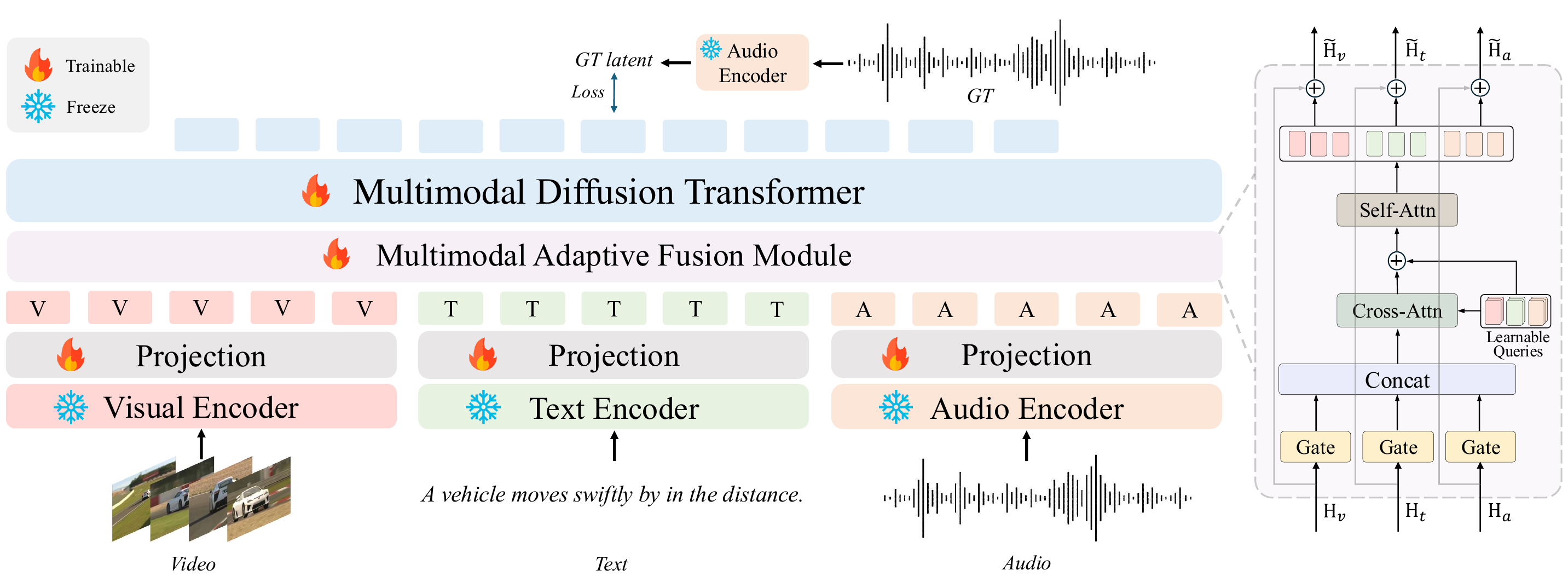}
    \caption{
        \textbf{The AudioX pretraining framework.}
        Specialized encoders process diverse modalities, and a MAF module unifies these signals into a conditioning embedding $H_c$.
        The MMDiT backbone processes the latent input $z_t$, conditioning on $H_c$ via cross-attention to generate high-quality audio and music.
        ($z_t$ and $H_c$ notations are omitted for visual clarity.)
    }
    \label{fig:framework}
\end{figure*}

\subsection{Model design}

The pretraining framework, as shown in Fig.~\ref{fig:framework}, is built upon a MMDiT backbone designed for high-fidelity audio synthesis. Given video $\mathbf{X}_{\texttt{v}}$, text $\mathbf{X}_{\texttt{t}}$, and audio $\mathbf{X}_{\texttt{a}}$, each modality is passed through corresponding specialized encoders. To capture the temporal dynamics, the resulting video and audio features are then processed by a temporal transformer. Finally, the features from all three modalities are mapped through a projection head to produce the domain-specific embeddings ($\mathbf{H}_{\texttt{v}}$, $\mathbf{H}_{\texttt{t}}$, $\mathbf{H}_{\texttt{a}}$). These embeddings are then fused into a unified condition embedding, which is ultimately passed to the MMDiT to guide the generation process.

A key challenge in training a unified model is that signals from different modalities can interfere with each other, making effective fusion and well-aligned conditioning critical. To address this, we introduce the lightweight Multimodal Adaptive Fusion (MAF) module. As shown in Fig.~\ref{fig:framework} (right), the MAF module operates as follows: \emph{First}, the initial feature embeddings from each modality are fed into \emph{gates}, which filter and reweight them to suppress noise and retain the most informative cues. \emph{Next}, the gated embeddings are concatenated and attended by \emph{learnable queries} via cross-attention. These queries are organized into three modality-specific sets, acting as experts to assess and aggregate evidence across the different data streams. \emph{Finally}, a \emph{self-attention} layer consolidates this aggregated context, and the refined information is dispatched back to the modality paths via residual updates.
This process yields calibrated, modality-specific outputs which are then concatenated to form the final multimodal condition embedding, $\mathbf{H}_{\texttt{c}}$:

\begin{equation}
\begin{aligned}
\tilde{\mathbf{H}}_{\texttt{v}},\ \tilde{\mathbf{H}}_{\texttt{t}},\ \tilde{\mathbf{H}}_{\texttt{a}}
&= \mathrm{MAF}\!\left(\mathbf{H}_{\texttt{v}},\,\mathbf{H}_{\texttt{t}},\,\mathbf{H}_{\texttt{a}}\right),\\
\mathbf{H}_{\texttt{c}}
&= \mathrm{Concat}\!\left(\tilde{\mathbf{H}}_{\texttt{v}},\,\tilde{\mathbf{H}}_{\texttt{t}},\,\tilde{\mathbf{H}}_{\texttt{a}}\right).
\end{aligned}
\end{equation}

This final embedding, along with a continuous timestep $t$, is what conditions the MMDiT backbone for the final audio synthesis. As we demonstrate in our ablation studies (Sec.~\ref{sec:ablation_study}), the MAF module is essential for reducing cross-modal interference while improving both the overall generation quality on multimodal tasks and the model's instruction-following capabilities.

\subsection{Training}
\label{sec:training}

The objective of the pretraining stage is to effectively integrate multimodal inputs and optimize the \modelbase\ teacher under a flow matching framework, producing high-quality audio or music conditioned on diverse multimodal inputs. The details of the training data are provided in Table~\ref{tab:dataset_overview} in the Appendix. During training, for each pair ($\mathbf{X}_{\texttt{v}}$, $\mathbf{X}_{\texttt{t}}$, $\mathbf{X}_{\texttt{a}}$ $;$ $\mathbf{A}$), where $\mathbf{A}$ is the ground truth we aim to generate, if the pair lacks video or audio modality input, we use zero-padding to fill the missing modality. If it lacks text modality input, we substitute with natural language descriptions, such as ``Generate music for the video.'' for the video-to-music generation task. For the tasks of audio inpainting and music completion, the audio modality input is required. In audio inpainting, $\mathbf{X}_{\texttt{a}}$ is a masked version of the ground truth audio $\mathbf{A}$, and the model's objective is to fill in the masked sections. For music completion, $\mathbf{X}_{\texttt{a}}$ is the preceding music segment of $\mathbf{A}$, and the model aims to generate the subsequent music segment of $\mathbf{X}_{\texttt{a}}$.

\noindent\textbf{Flow Matching process.} 
The MMDiT model processes the multimodal embedding $\mathbf{H}_{\texttt{c}}$ in the latent space through a flow matching denoising paradigm. Initially, the ground truth $\mathbf{A}$ is encoded using an encoder $\mathcal{E}$, which projects $\mathbf{A}$ into the latent space, yielding the target data representation $\mathbf{z}_0 = \mathcal{E}(\mathbf{A})$ at timestep $t=0$. The initial noise distribution is defined as a standard Gaussian $\mathbf{z}_1 \sim \mathcal{N}(0, \mathbf{I})$ at timestep $t=1$.

Instead of a Markov denoising Flow Matching constructs a continuous-time ordinary differential equation (ODE) to map the noise distribution to the data distribution. We adopt a simple straight-line path, where the intermediate latent state $\mathbf{z}_t$ at any timestep $t \in [0, 1]$ is defined as:
\begin{equation}
\mathbf{z}_t = t \mathbf{z}_1 + (1-t) \mathbf{z}_0
\end{equation}

The corresponding target vector field (velocity) that drives this transformation is simply the difference between the noise and the target data:
\begin{equation}
\mathbf{u}_t = \mathbf{z}_1 - \mathbf{z}_0
\end{equation}

\begin{figure*}[t]
    \centering
    \includegraphics[width=1\textwidth]{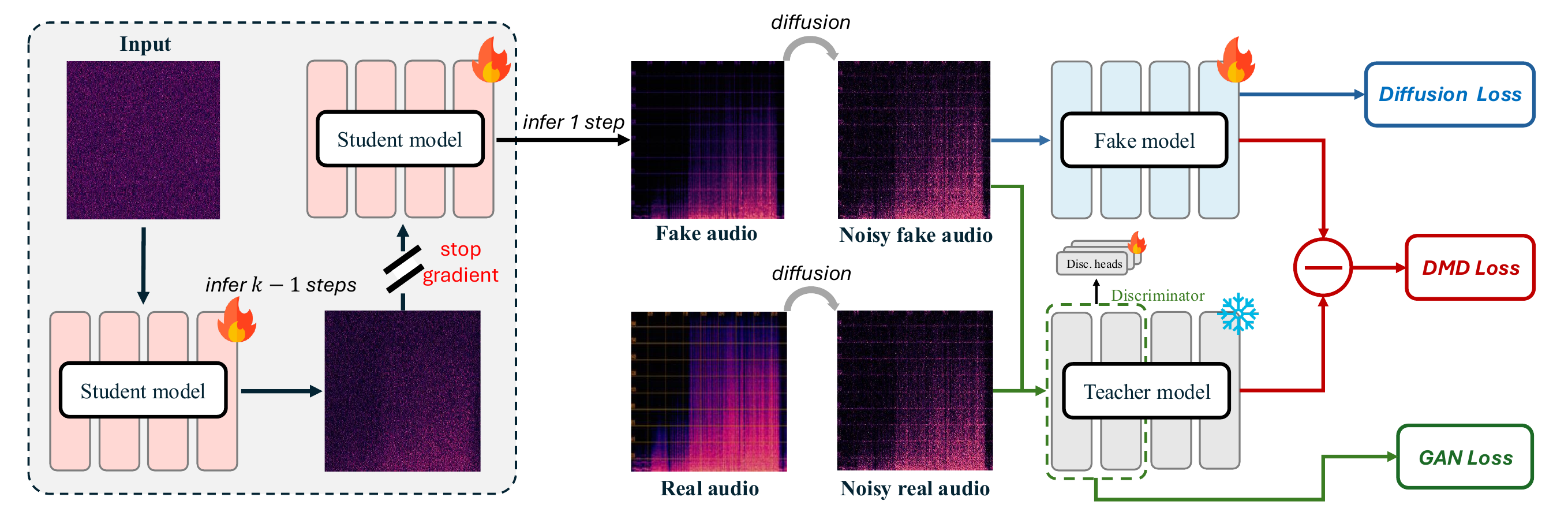}
    \caption{
        \textbf{The \model\ acceleration framework.}
        The generator is optimized with two objectives: a DMD loss derived from the discrepancy between the teacher and the fake model, and an adversarial loss from the diffusion-based discriminator. The auxiliary fake model is trained separately with a diffusion loss to fit the distribution of student-generated samples.
        Gradients are stopped through the rollout history and the frozen teacher branch. 
    }
    \label{fig:distillation_framework}
\end{figure*}

We train the MMDiT network, denoted as $v_\theta$, to predict this vector field. The model takes the intermediate state $\mathbf{z}_t$, timestep $t$, and the multimodal condition embedding $\mathbf{H}_{\texttt{c}}$ as inputs. The objective is to minimize the mean squared error between the predicted velocity and the target velocity:
\begin{equation}
\min _{\theta} \mathbb{E}_{t, \mathbf{z}_0, \mathbf{z}_1}\left\|v_\theta\left(\mathbf{z}_t, t, \mathbf{H}_{\texttt{c}}\right) - \left(\mathbf{z}_1 - \mathbf{z}_0\right)\right\|_2^2 .
\end{equation}

Training the MMDiT via this Flow Matching objective, we effectively unify multimodal inputs into the latent space, enabling the high-quality audio and music generation that is strictly aligned with the input conditions.

%% file: sec/5_Step_Distillation.tex
\section{Step Distillation}

Fig.~\ref{fig:distillation_framework} illustrates the overall distillation pipeline.
We aim to train an $N$-step student model and first construct a discrete timestep set $\mathcal{T}_N=\{t_N,\ldots,t_1,t_0\}$, where $t_N=1$ corresponds to pure noise and $t_0=0$ corresponds to clean data.
At each training iteration, we randomly sample one interval index $k\in\{1,\ldots,N\}$.
Starting from Gaussian noise, the student is rolled out along the preceding steps with a denoise-renoise paradigm: at each previous step, the student estimates a clean target, and the latent is then re-noised to the next scheduled timestep.
After the detached rollout reaches $\mathbf{z}_{t_k}$, the student performs the $k$-th denoising step with gradient enabled and predicts a clean estimate $\hat{\mathbf{z}}_0$, which represents a sample from the current student-induced distribution.
The student is optimized by a distribution matching objective, whose gradient is estimated using the real score provided by the frozen teacher and the fake score estimated by an auxiliary fake model.
The fake model is trained separately on stop-gradient student samples to track the evolving student-induced distribution.
In addition, an adversarial loss is applied to the student output to improve perceptual realism.

\subsection{Distribution Matching Distillation}

While the pretraining stage yields a model capable of high-fidelity synthesis,
solving the continuous-time ODE still requires many sequential function evaluations.
To enable real-time generation, we distill the pretrained teacher velocity field
$\mathbf{v}_{\theta}$ into an efficient student model $\mathbf{v}_{\phi}$.

The core principle of Distribution Matching Distillation (DMD)~\cite{dmd} is to
match the distribution induced by the student to the real data distribution. At a
sampled rollout step $t_k$, the student predicts a velocity
$\mathbf{v}_{\phi}(\mathbf{z}_{t_k}, t_k, \mathbf{H}_{\texttt{c}})$, from which we
estimate the clean sample as
\begin{equation}
\hat{\mathbf{z}}_0
=
\mathbf{z}_{t_k}
-
t_k\,\mathbf{v}_{\phi}(\mathbf{z}_{t_k}, t_k, \mathbf{H}_{\texttt{c}}).
\label{eq:student_clean_estimate}
\end{equation}
During inference, the next intermediate state is obtained by re-injecting noise to
this estimate, following the same denoise-renoise transition used in the rollout.

Directly characterizing the distribution induced by the student is intractable.
We therefore introduce an auxiliary fake model $\mathbf{v}_{\psi}$ to track the
velocity field of the current student distribution. Given the student prediction
$\hat{\mathbf{z}}_0$, we sample an evaluation timestep $\tau\in[0,1]$ and construct
a perturbed state by linear interpolation,
\begin{equation}
\mathbf{z}_{\tau}
=
\tau \mathbf{z}_{1} + (1-\tau)\hat{\mathbf{z}}_{0},
\qquad
\mathbf{z}_{1}\sim\mathcal{N}(\mathbf{0},\mathbf{I}).
\label{eq:dmd_interpolation}
\end{equation}
Both the frozen teacher $\mathbf{v}_{\theta}$ and the auxiliary fake model
$\mathbf{v}_{\psi}$ are evaluated at the same perturbed state and condition. We
define their velocity discrepancy as
\begin{equation}
\Delta\mathbf{v}_{\tau}
=
\mathbf{v}_{\theta}(\mathbf{z}_{\tau},\tau,\mathbf{H}_{\texttt{c}})
-
\mathbf{v}_{\psi}(\mathbf{z}_{\tau},\tau,\mathbf{H}_{\texttt{c}}).
\label{eq:vf_gap}
\end{equation}
We use this discrepancy as a velocity-space distribution-matching signal for
updating the student. During the student update, both $\mathbf{v}_{\theta}$ and
$\mathbf{v}_{\psi}$ are kept frozen; their outputs are used only to form
$\Delta\mathbf{v}_{\tau}$, and gradients are propagated only through the
student-dependent state $\mathbf{z}_{\tau}$. The resulting student-gradient
estimator is
\begin{equation}
\nabla_{\phi}\mathcal{L}_{\mathrm{DMD}}
=
\mathbb{E}_{\tau,\mathbf{z}_{1}}
\left[
\omega_{\tau}
\Delta\mathbf{v}_{\tau}^{\top}
\frac{\partial \mathbf{z}_{\tau}}{\partial \phi}
\right],
\label{eq:flow_dmd_gradient}
\end{equation}
where $\omega_{\tau}$ is a timestep-dependent weighting factor. Since
$\mathbf{z}_{\tau}$ depends on the student output $\hat{\mathbf{z}}_0$, this update
optimizes $\phi$ through the path
$\phi \rightarrow \hat{\mathbf{z}}_0 \rightarrow \mathbf{z}_{\tau}$.

The fake model is trained in a separate step to follow the evolving distribution
induced by the student. Given stop-gradient student samples
$\mathrm{sg}(\hat{\mathbf{z}}_0)$, we construct
\begin{equation}
\tilde{\mathbf{z}}_{\tau}
=
\tau\mathbf{z}_{1}
+
(1-\tau)\mathrm{sg}(\hat{\mathbf{z}}_0),
\label{eq:fake_interpolation}
\end{equation}
and optimize $\mathbf{v}_{\psi}$ with the standard flow matching objective
\begin{equation}
\mathcal{L}_{\mathrm{fake}}
=
\mathbb{E}_{\tau,\mathbf{z}_{1}}
\left[
\left\|
\mathbf{v}_{\psi}(\tilde{\mathbf{z}}_{\tau},\tau,\mathbf{H}_{\texttt{c}})
-
\left(\mathbf{z}_{1}-\mathrm{sg}(\hat{\mathbf{z}}_0)\right)
\right\|_2^2
\right].
\label{eq:fake_model_loss}
\end{equation}
This loss updates only the fake model, enabling it to track the current
student-induced distribution. The teacher branch remains frozen throughout
training and only supplies the reference velocity field.

\subsection{Diffusion-based Discriminator}

While the Distribution Matching Loss ($\mathcal{L}_{DM}$) rigorously aligns the distributions, it may occasionally fall short in capturing high-frequency acoustic textures and fine-grained perceptual details. To further enhance the realism and fidelity of the synthesized audio, we incorporate an adversarial generative training objective into the student model.

To construct a robust discriminator without the prohibitive computational cost of training from scratch, we leverage the deep, condition-aligned representations inherently captured by the pretrained teacher model. Specifically, we extract the first $L$ transformer blocks from the frozen teacher MMDiT $\mathbf{v}_\theta$ to serve as a feature extraction backbone. A lightweight discriminator head—composed of linear projection layers—is then appended on top of these blocks to predict the authenticity score. During the training process, the teacher backbone remains strictly frozen, and only the parameters of the discriminator head are updated. This design exploits the teacher's rich multimodal latent space as a powerful prior for discrimination.

Furthermore, following the practice in noisy-latent adversarial training, the discriminator operates on slightly perturbed latents rather than pristine clean outputs. This strategy prevents the discriminator from overfitting to low-level artifacts and provides more informative gradients. Specifically, we sample a small diffusion timestep $t_d \sim \mathcal{U}(0, 0.2)$ and inject noise into both the real clean data $\mathbf{z}_0$ and the student's estimated target $\hat{\mathbf{z}}_0$ to obtain their noisy counterparts, $\mathbf{z}_{t_d}$ and $\hat{\mathbf{z}}_{t_d}$, respectively. 

Let $D(\cdot, t_d, \mathbf{H}_{\texttt{c}})$ denote this diffusion-based discriminator. The discriminator is trained to distinguish between the real and implicitly generated noisy states using the standard hinge adversarial loss, corresponding to the GAN Loss in Fig.~\ref{fig:distillation_framework}:
\begin{equation}
\begin{aligned}
\mathcal{L}_{D} = \mathbb{E}_{\substack{\mathbf{z}_0, \mathbf{z}_1 \\ t_d, \mathbf{H}_{\texttt{c}}}}\Big[ &\max\big(0, 1 - D(\mathbf{z}_{t_d}, t_d, \mathbf{H}_{\texttt{c}})\big) \\
&+ \max\big(0, 1 + D(\hat{\mathbf{z}}_{t_d}, t_d, \mathbf{H}_{\texttt{c}})\big) \Big]
\end{aligned}
\end{equation}

Correspondingly, the student model $\mathbf{v}_\phi$ acts as the generator, with the objective of fooling the discriminator. The adversarial loss for updating the student model is formulated as:
\begin{equation}
\mathcal{L}_{adv} = - \mathbb{E}_{\substack{\mathbf{z}_1, t_k \\ t_d, \mathbf{H}_{\texttt{c}}}}\Big[D(\hat{\mathbf{z}}_{t_d}, t_d, \mathbf{H}_{\texttt{c}})\Big]
\end{equation}

By combining the distribution matching objective with the adversarial perceptual enhancement, the final overall training objective for the student model is defined as:
\begin{equation}
\mathcal{L}_{student} = \mathcal{L}_{DM} + \lambda_{adv} \mathcal{L}_{adv}
\end{equation}
where $\lambda_{adv}$ is a scalar hyperparameter controlling the weight of the adversarial loss.

%% file: sec/6_Experiment.tex
\section{Experiments}

In this section, we provide the implementation details of our experiments and conduct extensive evaluations. These assessments comprehensively measure the effectiveness of our proposed method from both subjective and objective viewpoints. The evaluations aim to offer valuable insights into the generation of audio and music from various inputs. Due to space constraints, additional results and analyses are provided in the Appendix.

\subsection{Implementation details}
\label{sec:implement_details}

\noindent\textbf{Pretraining Stage.} 
For encoding the visual features, we use CLIP-ViT-B/32~\citep{radford2021learning} to extract video frame features at a rate of 5 fps, and Synchformer~\citep{iashin2024synchformer} to extract synchronization features at 25 fps. The CLIP and Synchformer features are fused via addition. The text inputs are encoded using T5-base~\citep{raffel2020exploring}, while the audio is encoded and decoded using an audio Autoencoder~\citep{evans2024stable}. The model has a total of 2.7B parameters (2.4B trainable). Our proposed MAF module constitutes only 60M of these parameters, highlighting its lightweight nature. The MMDiT model, consisting of 24 layers, was trained from scratch without any pre-initialization. The training process uses the AdamW optimizer with a base learning rate of 1e-5, weight decay of 0.001, and a learning rate scheduler incorporating exponential ramp-up and decay phases. To improve inference stability, we maintain an exponential moving average (EMA) of the model weights. Training is conducted on three clusters of NVIDIA H800 GPUs, each with 80GB of memory. The total batch size is set to 240, and the model is trained for approximately 100k steps. Please refer to Appendix~\ref{sec:appendix_datasets} for further details on our training and evaluation datasets.

\noindent\textbf{Distillation Stage.} 
For the AudioX-Turbo distillation, we initialize both the student model $\mathbf{v}_\phi$ and the auxiliary fake model $\mathbf{v}_\psi$ with the weights of the fully converged pretrained teacher model. We compress the continuous ODE trajectory into a highly efficient $N=4$ step generation process, assigning uniform sampling probabilities across the discrete intervals. To eliminate the computational overhead of double forward passes during real-time inference, we explicitly bake Classifier-Free Guidance (CFG) into the student model. This is achieved by setting the teacher's guidance scale to 6.0 when generating the ground-truth deterministic endpoints. Furthermore, to ensure stable convergence and prevent the fake model from rapidly overfitting to the student's output, we employ an asymmetric update strategy: the student model is updated 5 times for every single update step of the fake model.

Regarding the adversarial and optimization configurations, the diffusion-based discriminator utilizes the frozen teacher's early transformer blocks as its backbone, operating on slightly perturbed latents with a noise level $t_d \sim \mathcal{U}(0, 0.2)$. The loss weights are strictly balanced, setting the distribution matching weight, fake model learning weight, and the adversarial loss weight ($\lambda_{adv}$) all to 1.0. Optimization is performed using the AdamW optimizer with a learning rate of 1e-5, $\beta=(0.9, 0.999)$, and a weight decay of 1e-3. We adopt an InverseLR scheduler with an inverse gamma of $10^6$, a power of 0.5, and a warm-up phase of 0.99 to dynamically adjust the learning rate. Similar to the pretraining stage, an EMA of the student's weights is continuously maintained to ensure high-fidelity inference.

\subsection{Evaluation metrics}
\label{sec:metrics}
To provide a comprehensive assessment of our model, we employ a suite of objective and subjective metrics. Further details for each metric are provided in the Appendix~\ref{sec:appendix_metrics}.
\paragraph{Objective Evaluation.} For overall audio quality and semantic alignment, we use several established metrics. These include: Kullback-Leibler Divergence (KL); Inception Score (IS); Fréchet Distance (FD) with PANNs embeddings \citep{kong2020panns}; Fréchet Audio Distance (FAD) with VGGish embeddings \citep{hershey2017cnn}; Production Complexity (PC) and Production Quality (PQ) \citep{tjandra2025meta}. As a prompt-free metric for both quality and diversity, we chose IS for the unified comparison in Fig.~\ref{fig:teaser}. For alignment, we use the CLAP score \citep{wu2023large} for text inputs and the Imagebind AV score \citep{girdhar2023imagebind} for video inputs.
To assess the model's instruction-following capabilities in T2A, we report metrics on two benchmarks. On our proposed \benchmark\ (detailed in Appendix \ref{sec:appendix_benchmark}), we measure category, count, ordering, and timestamp accuracy (Cat-acc, Cnt-acc, Ord-acc, TS-acc). On AudioTime \citep{xie2025audiotime}, we use its established metrics for Ordering, Duration, Frequency, and Timestamp. 
To assess inference efficiency, we report the number of function evaluations (NFE), the per-sample inference latency (Latency), and the real-time factor (RTF). For video-to-audio, we further report alignment accuracy (AlignAcc) \citep{luo2024diff} and audio-visual synchronization (AVSync) \citep{cheng2025mmaudio} to measure temporal correspondence between the generated audio and the input video.

\paragraph{Subjective Evaluation.} We conduct a formal user study with 10 professional audio experts to evaluate the subjective quality of our generated samples against baselines. The study follows prior work \citep{kreuk2022audiogen, liu2023audioldm}, where experts rate anonymized samples from 1 to 100 on Overall Quality (OVL) and Relevance (REL) to the prompt.

\subsection{Main results}
\label{sec:main_results}

This work introduces a unified model capable of generating audio and music from flexible combinations of video, text, and audio inputs. Through extensive experimentation, we benchmark our model against SOTA specialist models across all supported tasks. Results demonstrate that our single model consistently achieves SOTA or highly competitive performance on the majority of metrics.

\input{table/main_table}

\noindent\textbf{Audio generation.} Results of our audio generation are in Table~\ref{tab:main_table}, which includes the outcomes of generating audio or music from any combination of video and text modalities. The upper part of the table presents the audio generation tasks, while the lower part displays the music generation tasks.

For text-to-audio generation, we evaluate on the AudioCaps \citep{kim2019audiocaps} and VGGSound \citep{chen2020vggsound} datasets. On AudioCaps, our model achieves SOTA performance, while on VGGSound, the advantage is even more pronounced. This demonstrates that our model is a powerful text-to-audio generator. Furthermore, since VGGSound provides no native text captions, these text-to-audio results are obtained using captions generated by our annotation pipeline, demonstrating that our curated captions are reliable enough to support faithful T2A evaluation.
For video-to-audio generation, we evaluate on VGGSound \citep{chen2020vggsound} and benchmark against state-of-the-art video-conditioned specialist models~\citep{xing2024seeing,zhang2024foleycrafter, luo2024diff, wang2024frieren, cheng2025mmaudio}. Despite being a single unified model rather than a task-specific expert, \model\ delivers performance on par with these dedicated baselines.
For audio generation conditioned on both text and video, we benchmark against the strong baselines FoleyCrafter \citep{zhang2024foleycrafter} and MMAudio \citep{cheng2025mmaudio}, achieving results that are comparable to them. We find that when both text and video inputs are provided, the model can effectively generate better results.

The bottom part of Table~\ref{tab:main_table} shows the results of music generation tasks. On the V2M-bench~\citep{tian2025vidmuse}, we evaluate text-to-music, video-to-music, and video-and-text-to-music. The text-to-music task is additionally evaluated on the MusicCaps \citep{copet2024simple} dataset. Our model achieves SOTA performance across these tasks, demonstrating its effectiveness in generating high-quality music conditioned on diverse inputs.

\input{table/baselines_steps}

\noindent\textbf{Efficient inference.}
A central goal of \model\ is to retain the generation quality of multi-step diffusion models while drastically reducing the inference cost. To assess this, we evaluate \model, the multi-step teacher \modelbase, and a representative set of diffusion-based baselines on AudioCaps (T2A) and MusicCaps (T2M) under a range of sampling-step budgets $\{4, 50, 100, 200\}$. 
All baselines and \modelbase\ use CFG, doubling the cost per step ($\text{NFE}=2\times\text{steps}$), whereas \model\ distills CFG into the student and uses a single forward pass per step ($\text{NFE}=\text{steps}$). 
For a fair compute comparison, Table~\ref{tab:baselines_steps} reports NFE as a hardware-independent compute proxy, along with wall-clock latency and RTF measured under an identical protocol, where latency is averaged over 20 runs after 5 warm-ups.

As shown in Table~\ref{tab:baselines_steps}, using only $4$ NFE, \model\ matches or surpasses multi-step baselines that require up to $25\times$ more compute (e.g., $50$-step baselines at $100$ NFE), and this holds consistently on both AudioCaps and MusicCaps. 
Comparing \model\ against its teacher \modelbase\ further highlights the efficiency of our framework. At $4$ NFE, \model\ matches the multi-step \modelbase\ without noticeable degradation, while the baselines collapse when forced to $4$ steps. This shows that \model\ attains teacher-level quality at a small fraction of the inference cost, making high-quality anything-to-audio generation practical for low-latency applications.

\input{table/complex_tasks}

\noindent\textbf{Instruction-following text-to-audio generation.}
As shown in Figure~\ref{fig:teaser} and Table~\ref{tab:complex_tasks}, our models excel at tasks requiring fine-grained control. On \benchmark, both \modelbase\ and \model\ dominate the category, count, and ordering dimensions by large margins, more than doubling the best baseline on Cat-acc, Cnt-acc, and Ord-acc. On the timestamp dimension (TS-acc) they remain competitive with the strongest baselines. This advantage is reaffirmed on AudioTime, where our models achieve the best scores across all four metrics. Notably, the few-step \model\ stays on par with its multi-step teacher \modelbase\ and is even better on Ord-acc, indicating that distillation preserves fine-grained controllability rather than sacrificing it for speed. Collectively, these results underscore the superior fine-grained control of our framework.

\noindent\textbf{User study.} We conducted a user study to evaluate the quality of the generated audio and music. We randomly selected 25 samples for each audio generation task, including T2A, T2M, V2A, and V2M. 10 audio experts are asked to rate the quality of the generated audio and music. The results are shown in Fig.~\ref{fig:user_study} in the Appendix. The evaluation shows that our model achieves subjective SOTA performance in terms of OVL and REL scores in most tasks, indicating high user satisfaction.

To further demonstrate the versatility of our model, we present results for additional tasks, including audio inpainting, music completion, and image-to-audio generation, in Appendix~\ref{sec:appendix_comparisons}.
The results further underscore our model's strong performance and broad applicability across a variety of audio generation tasks.

\input{table/ablation_data_aug}

\subsection{Ablation study}

\label{sec:ablation_study}

In this section, we conduct a series of ablation studies to investigate the contribution of our key design choices. We systematically validate the efficacy of our data curation strategy and the architectural integrity of the proposed MAF module. An additional ablation study on the impact of different conditioning modalities is detailed in Appendix~\ref{sec:appendix_ablation}.

\noindent\textbf{Efficacy of data curation strategy.}
To verify the impact of our data curation strategy, we evaluate models trained on different textual supervision sources (Table~\ref{tab:ablation_data_aug}):
1) \texttt{Labels}: using raw class labels from the source datasets;
2) \texttt{AudioSetCaps}: using captions from a recent concurrent dataset \citep{bai2025audiosetcaps};
3) \texttt{QwenCap}: using captions generated directly by Qwen2-Audio;
4) \texttt{GeminiCap}: using only the initial annotations generated by Gemini 2.5 Pro; and
5) \texttt{GeminiCap-aug}: our full pipeline.
The results show that \texttt{GeminiCap-aug} outperforms all baselines, including the external AudioSetCaps dataset and the single-stage generation methods. It not only achieves the best scores on general-purpose tasks (T2A, V2A, TV2A) but also enhances the model's instruction-following capabilities. Collectively, these results validate the superior quality of our constructed dataset and the effectiveness of the proposed two-stage curation pipeline.
Notably, we observe that the benefits of high-quality textual supervision are not limited to text-to-audio generation. The marked improvement in the V2A task provides strong empirical evidence of a \textbf{\emph{cross-modal regularization effect}}. This insight leads to a crucial conclusion for future work: high-quality textual data should be viewed not only as an input, but also as an effective strategy for building more capable and robust multimodal models.

\input{table/ablation_maf}

\noindent\textbf{Architectural ablation of the MAF module.}
We conduct an architectural ablation of the MAF module to validate its design (Table~\ref{tab:maf_ablation}). The results confirm that each component is integral, with the most severe performance deterioration observed when the MAF module is omitted entirely. Removing the Gate mechanism or the Query-based attention individually also results in a performance decline, confirming their respective contributions. This analysis validates our design choices, underscoring that the complete MAF architecture is critical for optimal multimodal fusion, thereby enhancing cross-modal alignment and improving generation quality.

\noindent\textbf{Ablation study on efficient inference strategies.}
\input{table/ablation_acceleration}
We further analyze three design choices in the few-step distillation stage, with results summarized in Table~\ref{tab:acceleration_ablation}.
First, we vary the number of frozen teacher MMDiT blocks used as the discriminator backbone.
Using the first 6 blocks provides the best overall balance across AudioCaps and MusicCaps.
This suggests that relatively shallow teacher features already provide sufficient acoustic and condition-aligned evidence for discrimination, while deeper backbones may become overly semantic or overly strong, leading to less useful adversarial gradients for the student.
Second, we study the sampling probabilities over the four student timesteps.
The uniform schedule performs best, indicating that both high-noise stages, which shape global acoustic structure, and low-noise stages, which refine local timbre and temporal details, are important for few-step audio generation.
Biasing the training distribution toward either early or late timesteps degrades the overall balance between quality and distribution matching.
Finally, we isolate the adversarial objective.
The adversarial objective further improves perceptual fidelity and acoustic realism, as reflected by lower FAD/FD and higher IS.

\noindent\textbf{Study on training objective.}
\input{table/flow_vs_diffusion}
We compare the flow-matching objective with a standard diffusion objective under the same backbone and training budget. As shown in Table~\ref{tab:flow_vs_diffusion}, the two objectives achieve comparable generation quality, with only marginal differences across metrics on both benchmarks.
We nonetheless adopt flow matching for its compatibility with our distillation design: its velocity-field parameterization provides a convenient formulation of the distribution matching objective in Eq.~\ref{eq:flow_dmd_gradient}. This leads to a simpler and more direct training signal for the denoise-renoise rollout used to distill \modelbase\ into \model.

\noindent\textbf{Effect of the music training data.}
\input{table/music_data_ablation}
We further study the impact of scaling the video-music corpus by training on the original 360K subset versus the full V2M-500K.
As shown in Table~\ref{tab:music_data_ablation}, enlarging the corpus to 500K consistently improves performance across all metrics on music generation.
This confirms that the additional high-quality video-music pairs collected through our pipeline provide richer supervision and directly benefit downstream music generation.

\subsection{Discussion}
Our extensive experiments provide a multi-faceted validation of \model, consistently demonstrating state-of-the-art performance from broad audio generation to a commanding lead in fine-grained instruction following. Our ablation studies attribute this success to three complementary pillars. First, a data curation strategy that provides a rich semantic foundation through a cross-modal regularization effect. Second, an MAF architecture that translates these heterogeneous signals into precisely controlled outputs. Third, a distillation framework that compresses the multi-step teacher into a few-step student, preserving quality and controllability while drastically reducing inference cost. The synergy between this foundation, architecture, and distillation enables \model\ to unify generative versatility, fine-grained control, and practical efficiency within a single framework.

%% file: table/main_table.tex
\begin{table*}[!t]
    \centering
    \small
    \caption{
        \textbf{Performance evaluation across various tasks and datasets.} Task abbreviations are: T2A (Text-to-Audio), V2A (Video-to-Audio), TV2A (Text-and-Video-to-Audio), T2M (Text-to-Music), V2M (Video-to-Music), and TV2M (Text-and-Video-to-Music). For alignment (Align.), we use the CLAP score for text and the Imagebind AV score for video inputs.
    }
    \renewcommand{\arraystretch}{0.95}
    \adjustbox{max width=\textwidth}{
    \begin{tabular}{lll cc ccccc}
    \toprule

    Dataset & Method & Task & KL $\downarrow$ & IS $\uparrow$ & FD $\downarrow$ & FAD $\downarrow$ & PC $\uparrow$ & PQ $\uparrow$ & Align.$\uparrow$ \\
    \midrule
    \multirow{9}{*}{\centering\arraybackslash AudioCaps}
    & AudioGen~\citep{kreuk2022audiogen}       & T2A  & 1.39 & 10.22 & 13.29 & 1.72 & 3.26 & 5.25 & 0.27 \\
    & AudioLDM-L-Full~\citep{liu2023audioldm} & T2A  & 2.00 & 6.51 & 37.27 & 8.37 & 2.82 & 5.67 & 0.20 \\
    & AudioLDM-2-Large~\citep{liu2024audioldm} & T2A  & 1.49 & 8.46 & 26.34 & 1.97 & 2.86 & 5.77 & 0.22 \\
    & Tango 2~\citep{majumder2024tango}        & T2A  & \textbf{1.11} & 10.37 & \underline{12.22} & 3.20 & \textbf{3.63} & \underline{5.84} & \textbf{0.36} \\
    & Stable Audio Open~\citep{evans2024stable} & T2A  & 2.01 & 10.37 & 29.01 & 3.15 & 2.77 & \textbf{6.16} & 0.21 \\
    & MAGNET-large~\citep{DBLP:conf/iclr/ZivGLRKCDSA24} & T2A  & 1.62 & 7.46 & 24.88 & 2.99 & 3.25 & 5.15 & 0.15 \\
    & MMAudio~\citep{cheng2025mmaudio} & T2A  & 1.35 & 12.03 & 12.63 & 4.71 & 3.06 & 5.64 & \underline{0.30} \\
    & \cellcolor{cyan!7}\modelbase & \cellcolor{cyan!7}T2A & \cellcolor{cyan!7}\underline{1.29} & \cellcolor{cyan!7}\underline{12.46} & \cellcolor{cyan!7}\textbf{11.81} & \cellcolor{cyan!7}\textbf{1.65} & \cellcolor{cyan!7}3.20 & \cellcolor{cyan!7}5.65 & \cellcolor{cyan!7}0.27 \\
    & \cellcolor{cyan!7}\model & \cellcolor{cyan!7}T2A & \cellcolor{cyan!7}1.33 & \cellcolor{cyan!7}\textbf{12.37} & \cellcolor{cyan!7}12.29 & \cellcolor{cyan!7}\underline{1.68} & \cellcolor{cyan!7}\underline{3.50} & \cellcolor{cyan!7}5.65 & \cellcolor{cyan!7}0.29 \\
    \midrule
    \multirow{22}{*}{\centering\arraybackslash VGGSound}
    & AudioGen~\citep{kreuk2022audiogen}       & T2A  & 2.16 & 11.09 & 15.94 & 2.48 & 3.30 & 5.45 & 0.29 \\
    & AudioLDM-L-Full~\citep{liu2023audioldm} & T2A  & 2.41 & 6.52 & 31.15 & 7.05 & 2.93 & 5.99 & 0.27 \\
    & AudioLDM-2-Large~\citep{liu2024audioldm} & T2A  & 2.10 & 13.86 & 16.32 & 2.05 & 2.95 & \underline{6.35} & 0.30 \\
    & Tango 2~\citep{majumder2024tango}        & T2A  & 2.31 & 10.00 & 22.96 & 3.47 & \textbf{3.93} & 5.99 & 0.29 \\
    & Stable Audio Open~\citep{evans2024stable} & T2A  & 2.36 & 14.45 & 26.00 & 2.60 & 2.64 & \textbf{6.53} & \underline{0.33} \\
    & MAGNET-large~\citep{DBLP:conf/iclr/ZivGLRKCDSA24} & T2A  & \underline{2.03} & 8.53 & 22.17 & 2.74 & \underline{3.65} & 5.25 & 0.26 \\
    & MMAudio~\citep{cheng2025mmaudio} & T2A  & 2.17 & 17.83 & 11.52 & 2.50 & 3.02 & 6.12 & 0.32 \\
    & \cellcolor{cyan!7}\modelbase & \cellcolor{cyan!7}T2A & \cellcolor{cyan!7}2.06 & \cellcolor{cyan!7}\underline{19.12} & \cellcolor{cyan!7}\textbf{9.48} & \cellcolor{cyan!7}\underline{1.56} & \cellcolor{cyan!7}3.17 & \cellcolor{cyan!7}6.21 & \cellcolor{cyan!7}0.31 \\
    & \cellcolor{cyan!7}\model & \cellcolor{cyan!7}T2A & \cellcolor{cyan!7}\textbf{2.02} & \cellcolor{cyan!7}\textbf{19.72} & \cellcolor{cyan!7}\underline{10.12} & \cellcolor{cyan!7}\textbf{1.44} & \cellcolor{cyan!7}3.44 & \cellcolor{cyan!7}6.08 & \cellcolor{cyan!7}\textbf{0.35} \\
    \cmidrule(lr){2-10}

    & Seeing\&Hearing~\citep{xing2024seeing} & V2A  & 2.58 & 5.15 & 27.21 & 5.23 & 3.42 & 5.33 & \textbf{0.36} \\
    & FoleyCrafter~\citep{zhang2024foleycrafter}   & V2A  & 2.39 & 8.70 & 17.68 & 2.23 & 3.31 & 5.99 & 0.27 \\
    & Diff-Foley~\citep{luo2024diff}     & V2A  & 3.01 & 8.35 & 56.54 & 5.89 & 2.57 & 5.85 & 0.20 \\
    & FRIEREN~\citep{wang2024frieren}    & V2A  & 2.58 & 6.91 & 50.88 & 3.13 & 2.98 & 6.06 & 0.20 \\
    & VATT~\citep{liu2024tell}           & V2A  & \textbf{1.40} & 10.02 & 11.71 & 2.55 & 3.64 & 5.85 & 0.25 \\
    & VAB~\citep{su2024vision}           & V2A  & 2.30 & 8.15 & 20.21 & 3.05 & 3.52 & 5.93 & 0.24 \\
    & MMAudio~\citep{cheng2025mmaudio} & V2A  & \underline{1.97} & \textbf{14.95} & \textbf{6.18} & 2.04 & 3.38 & 5.91 & \underline{0.35} \\
    & \cellcolor{cyan!7}\modelbase & \cellcolor{cyan!7}V2A & \cellcolor{cyan!7}1.98 & \cellcolor{cyan!7}12.35 & \cellcolor{cyan!7}\underline{6.94} & \cellcolor{cyan!7}\underline{2.00} & \cellcolor{cyan!7}\underline{3.75} & \cellcolor{cyan!7}\textbf{6.31} & \cellcolor{cyan!7}0.28 \\
    & \cellcolor{cyan!7}\model & \cellcolor{cyan!7}V2A & \cellcolor{cyan!7}1.99 & \cellcolor{cyan!7}\underline{12.39} & \cellcolor{cyan!7}7.88 & \cellcolor{cyan!7}\textbf{1.34} & \cellcolor{cyan!7}\textbf{3.78} & \cellcolor{cyan!7}\underline{6.23} & \cellcolor{cyan!7}0.29 \\
    \cmidrule(lr){2-10}

    & FoleyCrafter~\citep{zhang2024foleycrafter}   & TV2A & 1.94 & 11.32 & 19.16 & 2.13 & 3.38 & 6.06 & 0.26 \\
    & MMAudio~\citep{cheng2025mmaudio} & TV2A  & \textbf{1.51} & \underline{17.79} & 6.60 & 2.20 & 3.31 & 5.99 & \textbf{0.33} \\
    & \cellcolor{cyan!7}\modelbase & \cellcolor{cyan!7}TV2A & \cellcolor{cyan!7}\underline{1.56} & \cellcolor{cyan!7}\textbf{17.81} & \cellcolor{cyan!7}\underline{6.59} & \cellcolor{cyan!7}\underline{1.86} & \cellcolor{cyan!7}\underline{3.66} & \cellcolor{cyan!7}\textbf{6.34} & \cellcolor{cyan!7}0.28 \\
    & \cellcolor{cyan!7}\model & \cellcolor{cyan!7}TV2A & \cellcolor{cyan!7}1.60 & \cellcolor{cyan!7}17.47 & \cellcolor{cyan!7}\textbf{5.78} & \cellcolor{cyan!7}\textbf{0.97} & \cellcolor{cyan!7}\textbf{3.67} & \cellcolor{cyan!7}\underline{6.25} & \cellcolor{cyan!7}\underline{0.29} \\
    \midrule
    \multirow{8}{*}{\centering\arraybackslash MusicCaps}
    & MusicGen~\citep{copet2024simple}       & T2M  & 1.43 & 2.24 & 25.40 & 4.55 & 5.19 & \underline{7.16} & 0.18 \\
    & AudioLDM-L-Full~\citep{liu2023audioldm} & T2M  & 1.45 & 2.49 & 34.44 & 6.34 & 4.72 & 6.10 & \underline{0.22} \\
    & AudioLDM-2-Large~\citep{liu2024audioldm} & T2M  & \underline{1.26} & 2.84 & 15.61 & 2.80 & 5.22 & 6.70 & \textbf{0.23} \\
    & TangoMusic~\citep{ghosal2023text} & T2M  & \textbf{1.13} & 2.86 & 15.00 & 1.88 & \underline{5.57} & 7.06 & \textbf{0.23} \\
    & Stable Audio Open~\citep{evans2024stable}  & T2M  & 1.51 & 2.94 & 36.33 & 3.23 & 3.91 & \textbf{7.18} & \textbf{0.23} \\
    & MAGNET-large~\citep{DBLP:conf/iclr/ZivGLRKCDSA24} & T2M  & 1.32 & 1.98 & 23.88 & 4.24 & \textbf{5.84} & 6.71 & 0.19 \\
    & \cellcolor{cyan!7}\modelbase & \cellcolor{cyan!7}T2M & \cellcolor{cyan!7}1.38 & \cellcolor{cyan!7}\underline{3.47} & \cellcolor{cyan!7}\underline{12.89} & \cellcolor{cyan!7}\underline{1.67} & \cellcolor{cyan!7}4.57 & \cellcolor{cyan!7}6.45 & \cellcolor{cyan!7}0.20 \\
    & \cellcolor{cyan!7}\model & \cellcolor{cyan!7}T2M & \cellcolor{cyan!7}1.31 & \cellcolor{cyan!7}\textbf{3.61} & \cellcolor{cyan!7}\textbf{9.50} & \cellcolor{cyan!7}\textbf{1.54} & \cellcolor{cyan!7}4.89 & \cellcolor{cyan!7}6.55 & \cellcolor{cyan!7}\underline{0.22} \\
    \midrule
    \multirow{16}{*}{\centering\arraybackslash V2M-bench}
    & MusicGen~\citep{copet2024simple}       & T2M  & 0.76 & 1.31 & 40.59 & 3.25 & 5.57 & 7.43 & 0.14 \\
    & AudioLDM-L-Full~\citep{liu2023audioldm} & T2M  & 0.72 & 1.37 & 36.63 & 2.97 & 5.08 & 7.01 & \underline{0.16} \\
    & AudioLDM-2-Large~\citep{liu2024audioldm} & T2M  & \underline{0.62} & 1.46 & 25.80 & \textbf{1.63} & 5.57 & 6.90 & 0.14 \\
    & TangoMusic~\citep{ghosal2023text}    & T2M  & 0.72 & 1.46 & 38.19 & 2.43 & \underline{5.78} & \underline{7.46} & 0.14 \\
    & Stable Audio Open~\citep{evans2024stable} & T2M  & 0.72 & 1.34 & 42.02 & 2.72 & 4.36 & \textbf{7.72} & \textbf{0.17} \\
    & MAGNET-large~\citep{DBLP:conf/iclr/ZivGLRKCDSA24} & T2M  & \textbf{0.60} & 1.26 & 34.24 & 3.15 & \textbf{5.89} & 7.04 & \textbf{0.17} \\
    & \cellcolor{cyan!7}\modelbase & \cellcolor{cyan!7}T2M & \cellcolor{cyan!7}0.71 & \cellcolor{cyan!7}\textbf{2.24} & \cellcolor{cyan!7}\textbf{11.06} & \cellcolor{cyan!7}\underline{2.13} & \cellcolor{cyan!7}5.28 & \cellcolor{cyan!7}7.10 & \cellcolor{cyan!7}0.13 \\
    & \cellcolor{cyan!7}\model & \cellcolor{cyan!7}T2M & \cellcolor{cyan!7}0.72 & \cellcolor{cyan!7}\underline{2.22} & \cellcolor{cyan!7}\underline{11.75} & \cellcolor{cyan!7}2.62 & \cellcolor{cyan!7}5.53 & \cellcolor{cyan!7}6.95 & \cellcolor{cyan!7}0.14 \\
    \cmidrule(lr){2-10}
    & Video2Music~\citep{kang2024video2music} & V2M & 1.78 & 1.01 & 144.88 & 18.72 & 3.34 & \underline{8.14} & 0.14 \\
    & MuMu-LLaMA~\citep{liu2024mumu} & V2M & 1.00 & 1.25 & 52.25 & 5.10 & \underline{5.60} & 7.97 & 0.18 \\
    & CMT~\citep{di2021video} & V2M & 1.22 & 1.24 & 85.70 & 8.64 & 4.98 & \textbf{8.20} & 0.12 \\
    & VidMuse~\citep{tian2025vidmuse} & V2M & 0.73 & 1.32 & 29.95 & 2.46 & \textbf{5.88} & 6.89 & \underline{0.20} \\
    & \cellcolor{cyan!7}\modelbase & \cellcolor{cyan!7}V2M & \cellcolor{cyan!7}\underline{0.49} & \cellcolor{cyan!7}\textbf{1.48} & \cellcolor{cyan!7}\underline{23.47} & \cellcolor{cyan!7}\textbf{1.49} & \cellcolor{cyan!7}5.22 & \cellcolor{cyan!7}7.27 & \cellcolor{cyan!7}\textbf{0.23} \\
    & \cellcolor{cyan!7}\model & \cellcolor{cyan!7}V2M & \cellcolor{cyan!7}\textbf{0.44} & \cellcolor{cyan!7}\underline{1.47} & \cellcolor{cyan!7}\textbf{19.66} & \cellcolor{cyan!7}\underline{1.56} & \cellcolor{cyan!7}5.46 & \cellcolor{cyan!7}7.23 & \cellcolor{cyan!7}\textbf{0.23} \\
    \cmidrule(lr){2-10}
    & \cellcolor{cyan!7}\modelbase & \cellcolor{cyan!7}TV2M & \cellcolor{cyan!7}\underline{0.49} & \cellcolor{cyan!7}\textbf{1.45} & \cellcolor{cyan!7}\underline{23.69} & \cellcolor{cyan!7}\textbf{1.50} & \cellcolor{cyan!7}\underline{5.20} & \cellcolor{cyan!7}\textbf{7.25} & \cellcolor{cyan!7}\underline{0.22} \\
    & \cellcolor{cyan!7}\model & \cellcolor{cyan!7}TV2M & \cellcolor{cyan!7}\textbf{0.43} & \cellcolor{cyan!7}\underline{1.44} & \cellcolor{cyan!7}\textbf{19.86} & \cellcolor{cyan!7}\underline{1.56} & \cellcolor{cyan!7}\textbf{5.45} & \cellcolor{cyan!7}\underline{7.22} & \cellcolor{cyan!7}\textbf{0.23} \\

    \bottomrule

    \end{tabular}
    }
    \label{tab:main_table}
\end{table*}

%% file: table/baselines_steps.tex
\begin{table*}[t]
    \centering
    \small
    \caption{
        \textbf{Performance vs.\ sampling steps.} We report the number of function evaluations (NFE) as a hardware-independent compute proxy, together with wall-clock latency and the real-time factor (RTF). CFG-based methods (all baselines and \modelbase) cost $\text{NFE}=2\times\text{Steps}$, whereas the CFG-free \model\ costs $\text{NFE}=\text{Steps}$. Latency is measured end-to-end on a single NVIDIA RTX 4090 GPU at batch size~1 for a 10-second clip (mean$\pm$std over 20 runs after 5 warm-ups), and $\text{RTF}=\text{latency}/\text{duration}$ ($<\!1$ is faster than real-time). 
        Best per column in \textbf{bold}, second best \underline{underlined}; cyan rows are ours.
    }
    \renewcommand{\arraystretch}{0.9}
    \adjustbox{max width=\textwidth}{
    \begin{tabular}{ll cccc | ccccccc}
    \toprule
    Dataset & Method & Steps & NFE & Latency (s) $\downarrow$ & RTF $\downarrow$ & KL $\downarrow$ & IS $\uparrow$ & FD $\downarrow$ & FAD $\downarrow$ & PC $\uparrow$ & PQ $\uparrow$ & Align.$\uparrow$ \\
    \midrule

    \multirow{29}{*}{\centering\arraybackslash AudioCaps}
    & \multirow{4}{*}{AudioLDM~\citep{liu2023audioldm}}
        &   4 &   8 &  \textbf{0.11}$\pm$0.010 & \textbf{0.01} & 2.65 &  3.87 & 59.57 & 18.93 & 2.68 & 5.55 & 0.12 \\
    & &  50 & 100 &  0.96$\pm$0.060 & 0.10 & 2.02 &  6.32 & 37.88 &  8.59 & 2.78 & 5.66 & 0.14 \\
    & & 100 & 200 &  1.85$\pm$0.036 & 0.19 & 2.16 &  6.41 & 37.86 &  8.43 & 2.81 & 5.77 & 0.15 \\
    & & 200 & 400 &  3.67$\pm$0.094 & 0.37 & 1.96 &  6.54 & 37.04 &  8.29 & 2.83 & 5.68 & 0.15 \\
    \cmidrule(lr){2-13}
    & \multirow{4}{*}{AudioLDM-2~\citep{liu2024audioldm}}
        &   4 &   8 &  0.39$\pm$0.065 & 0.04 & 2.40 &  4.42 & 47.97 & 10.28 & 2.88 & 5.58 & 0.09 \\
    & &  50 & 100 &  3.30$\pm$0.158 & 0.33 & 1.51 &  8.55 & 27.38 &  1.95 & 2.84 & 5.77 & 0.26 \\
    & & 100 & 200 &  6.61$\pm$0.333 & 0.66 & 1.59 &  8.41 & 26.99 &  2.08 & 2.87 & 5.77 & 0.27 \\
    & & 200 & 400 & 13.36$\pm$0.610 & 1.34 & 1.39 &  8.43 & 26.13 &  1.92 & 2.86 & 5.78 & 0.26 \\
    \cmidrule(lr){2-13}
    & \multirow{4}{*}{Stable Audio Open~\citep{evans2024stable}}
        &   4 &   8 &  1.33$\pm$0.071 & 0.13 & 3.36 &  5.46 & 62.04 & 12.87 & 2.53 & \underline{6.02} & 0.08 \\
    & &  50 & 100 & 10.48$\pm$0.106 & 1.05 & 2.12 & 10.21 & 30.22 &  3.28 & 2.70 & \textbf{6.12} & 0.02 \\
    & & 100 & 200 & 20.39$\pm$0.176 & 2.04 & 1.91 & 10.48 & 29.43 &  3.18 & 2.70 & \textbf{6.12} & 0.04 \\
    & & 200 & 400 & 40.06$\pm$0.214 & 4.01 & 1.93 & 10.36 & 28.74 &  3.05 & 2.70 & \textbf{6.12} & 0.04 \\
    \cmidrule(lr){2-13}
    & \multirow{4}{*}{Tango 2~\citep{majumder2024tango}}
        &   4 &   8 &  0.56$\pm$0.053 & 0.06 & 1.65 &  4.18 & 48.20 & 20.34 & 3.03 & 5.14 & 0.18 \\
    & &  50 & 100 &  5.76$\pm$0.086 & 0.58 & 1.19 & 10.12 & 12.89 &  3.38 & 3.55 & 5.81 & \textbf{0.36} \\
    & & 100 & 200 & 11.47$\pm$0.087 & 1.15 & \underline{1.11} & 10.21 & 12.44 &  3.47 & \textbf{3.61} & 5.84 & \textbf{0.36} \\
    & & 200 & 400 & 22.74$\pm$0.187 & 2.27 & \textbf{1.10} & 10.41 & \underline{12.13} &  3.07 & \underline{3.60} & 5.85 & \underline{0.35} \\
    \cmidrule(lr){2-13}
    & \multirow{4}{*}{MMAudio~\citep{cheng2025mmaudio}}
        &   4 &   8 &  0.62$\pm$0.063 & 0.06 & 2.33 &  4.28 & 49.72 & 10.81 & 2.45 & 4.39 & 0.08 \\
    & &  50 & 100 &  2.18$\pm$0.074 & 0.22 & 1.45 & 11.92 & 13.03 &  4.84 & 2.99 & 5.66 & 0.21 \\
    & & 100 & 200 &  4.04$\pm$0.195 & 0.40 & 1.37 & 12.06 & 12.86 &  4.93 & 3.01 & 5.67 & 0.21 \\
    & & 200 & 400 &  7.27$\pm$0.103 & 0.73 & 1.35 & 11.84 & 12.49 &  4.59 & 3.02 & 5.65 & 0.21 \\
    \cmidrule(lr){2-13}
    & \cellcolor{cyan!7} & \cellcolor{cyan!7}  4 & \cellcolor{cyan!7}  8 & \cellcolor{cyan!7} 0.46$\pm$0.038 & \cellcolor{cyan!7}0.05 & \cellcolor{cyan!7}4.31 & \cellcolor{cyan!7} 3.36 & \cellcolor{cyan!7}77.79 & \cellcolor{cyan!7}16.80 & \cellcolor{cyan!7}2.89 & \cellcolor{cyan!7}5.15 & \cellcolor{cyan!7}0.04 \\
    & \cellcolor{cyan!7} & \cellcolor{cyan!7} 50 & \cellcolor{cyan!7}100 & \cellcolor{cyan!7} 1.72$\pm$0.126 & \cellcolor{cyan!7}0.17 & \cellcolor{cyan!7}1.29 & \cellcolor{cyan!7}12.47 & \cellcolor{cyan!7}12.58 & \cellcolor{cyan!7} 1.69 & \cellcolor{cyan!7}3.16 & \cellcolor{cyan!7}5.80 & \cellcolor{cyan!7}0.29 \\
    & \cellcolor{cyan!7} & \cellcolor{cyan!7}100 & \cellcolor{cyan!7}200 & \cellcolor{cyan!7} 2.98$\pm$0.121 & \cellcolor{cyan!7}0.30 & \cellcolor{cyan!7}1.29 & \cellcolor{cyan!7}\underline{12.48} & \cellcolor{cyan!7}12.30 & \cellcolor{cyan!7} \underline{1.68} & \cellcolor{cyan!7}3.16 & \cellcolor{cyan!7}5.81 & \cellcolor{cyan!7}0.29 \\
    & \cellcolor{cyan!7}\multirow{-4}{*}{\modelbase} & \cellcolor{cyan!7}200 & \cellcolor{cyan!7}400 & \cellcolor{cyan!7} 5.49$\pm$0.119 & \cellcolor{cyan!7}0.55 & \cellcolor{cyan!7}1.29 & \cellcolor{cyan!7}\textbf{12.51} & \cellcolor{cyan!7}\textbf{11.98} & \cellcolor{cyan!7} \textbf{1.57} & \cellcolor{cyan!7}3.16 & \cellcolor{cyan!7}5.81 & \cellcolor{cyan!7}0.29 \\
    \cmidrule(lr){2-13}
    & \cellcolor{cyan!7}\model & \cellcolor{cyan!7}4 & \cellcolor{cyan!7}4 & \cellcolor{cyan!7}\underline{0.24}$\pm$0.002 & \cellcolor{cyan!7}\underline{0.02}
        & \cellcolor{cyan!7}1.33 & \cellcolor{cyan!7}12.37 & \cellcolor{cyan!7}12.29 & \cellcolor{cyan!7}\underline{1.68}
        & \cellcolor{cyan!7}3.50 & \cellcolor{cyan!7}5.65 & \cellcolor{cyan!7}0.29 \\

    \midrule

    \multirow{21}{*}{\centering\arraybackslash MusicCaps}
    & \multirow{4}{*}{TangoMusic~\citep{ghosal2023text}}
        &   4 &   8 &  0.57$\pm$0.055 & 0.06 & 2.29 &  1.97 & 75.46 & 27.72 & 3.61 & 5.28 & 0.06 \\
    & &  50 & 100 &  5.78$\pm$0.066 & 0.58 & 1.23 &  2.73 & 15.45 &  1.90 & \underline{5.55} & 7.01 & 0.23 \\
    & & 100 & 200 & 11.32$\pm$0.156 & 1.13 & \textbf{1.12} &  2.78 & 15.11 &  2.07 & \textbf{5.56} & 7.06 & 0.23 \\
    & & 200 & 400 & 22.64$\pm$0.226 & 2.26 & \textbf{1.12} &  2.85 & 14.97 &  1.86 & \textbf{5.56} & 7.08 & 0.23 \\
    \cmidrule(lr){2-13}
    & \multirow{4}{*}{AudioLDM~\citep{liu2023audioldm}}
        &   4 &   8 &  \textbf{0.11}$\pm$0.012 & \textbf{0.01} & 1.76 &  2.05 & 52.14 & 13.09 & 4.36 & 5.85 & 0.16 \\
    & &  50 & 100 &  0.93$\pm$0.060 & 0.09 & 1.54 &  2.40 & 35.14 &  6.46 & 4.71 & 5.97 & 0.23 \\
    & & 100 & 200 &  1.83$\pm$0.056 & 0.18 & 1.44 &  2.41 & 34.77 &  6.56 & 4.74 & 6.17 & 0.23 \\
    & & 200 & 400 &  3.70$\pm$0.116 & 0.37 & 1.44 &  2.51 & 34.00 &  6.23 & 4.75 & 6.15 & 0.23 \\
    \cmidrule(lr){2-13}
    & \multirow{4}{*}{AudioLDM-2~\citep{liu2024audioldm}}
        &   4 &   8 &  0.40$\pm$0.083 & 0.04 & 1.61 &  2.27 & 35.89 &  5.88 & 4.97 & 6.12 & 0.13 \\
    & &  50 & 100 &  3.42$\pm$0.233 & 0.34 & 1.33 &  2.85 & 16.14 &  2.80 & 5.13 & 6.56 & 0.23 \\
    & & 100 & 200 &  6.83$\pm$0.328 & 0.68 & 1.25 &  2.89 & 15.65 &  2.96 & 5.14 & 6.63 & 0.23 \\
    & & 200 & 400 & 13.86$\pm$0.796 & 1.39 & 1.16 &  2.81 & 15.05 &  2.77 & 5.22 & 6.72 & 0.23 \\
    \cmidrule(lr){2-13}
    & \multirow{4}{*}{Stable Audio Open~\citep{evans2024stable}}
        &   4 &   8 &  1.33$\pm$0.070 & 0.13 & 3.09 &  2.35 & 101.74 & 20.32 & 1.97 & 6.64 & 0.01 \\
    & &  50 & 100 & 10.43$\pm$0.105 & 1.04 & 1.55 &  2.83 & 37.08 &  3.37 & 3.83 & \underline{7.18} & 0.22 \\
    & & 100 & 200 & 20.32$\pm$0.137 & 2.03 & 1.44 &  3.07 & 36.39 &  3.44 & 3.95 & \underline{7.18} & 0.22 \\
    & & 200 & 400 & 40.19$\pm$0.209 & 4.02 & 1.43 &  2.91 & 35.66 &  3.14 & 3.97 & \textbf{7.20} & 0.22 \\
    \cmidrule(lr){2-13}
    & \cellcolor{cyan!7} & \cellcolor{cyan!7}  4 & \cellcolor{cyan!7}  8 & \cellcolor{cyan!7} 0.46$\pm$0.050 & \cellcolor{cyan!7}0.05 & \cellcolor{cyan!7}2.34 & \cellcolor{cyan!7}2.61 & \cellcolor{cyan!7}51.86 & \cellcolor{cyan!7}6.74 & \cellcolor{cyan!7}3.74 & \cellcolor{cyan!7}5.88 & \cellcolor{cyan!7}0.04 \\
    & \cellcolor{cyan!7} & \cellcolor{cyan!7} 50 & \cellcolor{cyan!7}100 & \cellcolor{cyan!7} 1.68$\pm$0.057 & \cellcolor{cyan!7}0.17 & \cellcolor{cyan!7}1.40 & \cellcolor{cyan!7}3.45 & \cellcolor{cyan!7} 12.61 & \cellcolor{cyan!7}1.72 & \cellcolor{cyan!7}4.82 & \cellcolor{cyan!7}6.57 & \cellcolor{cyan!7}\textbf{0.24} \\
    & \cellcolor{cyan!7} & \cellcolor{cyan!7}100 & \cellcolor{cyan!7}200 & \cellcolor{cyan!7} 2.95$\pm$0.113 & \cellcolor{cyan!7}0.29 & \cellcolor{cyan!7}1.38 & \cellcolor{cyan!7}\textbf{3.67} & \cellcolor{cyan!7} \underline{12.58} & \cellcolor{cyan!7}\underline{1.69} & \cellcolor{cyan!7}4.80 & \cellcolor{cyan!7}6.58 & \cellcolor{cyan!7}0.23 \\
    & \cellcolor{cyan!7}\multirow{-4}{*}{\modelbase} & \cellcolor{cyan!7}200 & \cellcolor{cyan!7}400 & \cellcolor{cyan!7} 5.57$\pm$0.126 & \cellcolor{cyan!7}0.56 & \cellcolor{cyan!7}1.38 & \cellcolor{cyan!7}\underline{3.62} & \cellcolor{cyan!7} 12.62 & \cellcolor{cyan!7}\underline{1.69} & \cellcolor{cyan!7}4.80 & \cellcolor{cyan!7}6.60 & \cellcolor{cyan!7}\textbf{0.24} \\
    \cmidrule(lr){2-13}
    & \cellcolor{cyan!7}\model & \cellcolor{cyan!7}4 & \cellcolor{cyan!7}4 & \cellcolor{cyan!7}\underline{0.24}$\pm$0.002 & \cellcolor{cyan!7}\underline{0.02}
        & \cellcolor{cyan!7}\underline{1.31} & \cellcolor{cyan!7}3.61 & \cellcolor{cyan!7}\textbf{9.50} & \cellcolor{cyan!7}\textbf{1.54}
        & \cellcolor{cyan!7}4.89 & \cellcolor{cyan!7}6.55 & \cellcolor{cyan!7}0.22 \\

    \bottomrule
    \end{tabular}
    }
    \label{tab:baselines_steps}
\end{table*}

%% file: table/complex_tasks.tex
\begin{table*}[h]
    \centering
    \small
    \caption{
        \textbf{Evaluation of instruction-following T2A ability on \benchmark\ and AudioTime.} Best per column is in \textbf{bold}, second best \underline{underlined}; cyan rows mark our methods.
    }
    \renewcommand{\arraystretch}{1}
    \adjustbox{max width=\textwidth}{
    \begin{tabular}{l cccc cccc}
    \toprule
    \multirow{2}{*}{Method} & \multicolumn{4}{c}{\benchmark} & \multicolumn{4}{c}{AudioTime} \\
    \cmidrule(lr){2-5} \cmidrule(lr){6-9}
    & Cat-acc $\uparrow$ & Cnt-acc $\uparrow$ & Ord-acc $\uparrow$ & TS-acc $\uparrow$ & Ordering $\downarrow$ & Duration $\downarrow$ & Frequency $\downarrow$ & Timestamp $\uparrow$ \\
    \midrule
    AudioGen~\citep{kreuk2022audiogen}            & 24.40 &  5.40 &  6.00 & 18.40 & 0.91 & 3.73 & 1.58 & 0.54 \\
    AudioLDM~\citep{liu2023audioldm}              & 18.60 &  4.00 &  3.40 & 11.60 & 0.97 & 3.41 & 1.54 & 0.41 \\
    AudioLDM-2~\citep{liu2024audioldm}            & 20.10 & 7.40 &  1.20 & 13.40 & 0.96 & 3.40 & 1.64 & 0.54 \\
    Tango 2~\citep{majumder2024tango}             & 25.20 &  4.60 & 10.20 & 18.80 & 0.86 & 3.70 & 1.52 & 0.61 \\
    Make-An-Audio 2~\citep{huang2023make}         & 32.40 &  4.00 & 19.80 & 18.80 & 0.76 & 3.40 & 1.42 & 0.56 \\
    Stable Audio Open~\citep{evans2024stable}     & 31.20 &  9.80 &  6.00 & \textbf{21.80} & 0.98 & 3.07 & 1.46 & 0.53 \\
    MMAudio~\citep{cheng2025mmaudio}              & 26.60 &  4.80 &  2.40 & \underline{21.40} & 0.98 & 3.33 & 1.54 & 0.50 \\
    \cmidrule(lr){1-9}
    \cellcolor{cyan!7}\modelbase & \cellcolor{cyan!7}\textbf{75.00} & \cellcolor{cyan!7}\textbf{24.00} & \cellcolor{cyan!7}\underline{52.80} & \cellcolor{cyan!7}17.40 & \cellcolor{cyan!7}\textbf{0.58} & \cellcolor{cyan!7}\textbf{1.71} & \cellcolor{cyan!7}\textbf{0.89} & \cellcolor{cyan!7}\textbf{0.70} \\
    \cellcolor{cyan!7}\model     & \cellcolor{cyan!7}\underline{74.80} & \cellcolor{cyan!7}\underline{21.80} & \cellcolor{cyan!7}\textbf{55.40} & \cellcolor{cyan!7}18.80 & \cellcolor{cyan!7}\underline{0.63} & \cellcolor{cyan!7}\underline{1.80} & \cellcolor{cyan!7}\underline{0.96} & \cellcolor{cyan!7}\underline{0.66} \\
    \bottomrule
    \end{tabular}}
    \label{tab:complex_tasks}
\end{table*}

%% file: table/ablation_data_aug.tex
\begin{table*}[h]
    \centering
    \small
    \caption{ 
 \textbf{Ablation study on data curation strategies.}
    We compare our model's performance when trained with captions from different sources.
    The results show a clear trend of improvement with higher-quality data.
    Our full pipeline (\texttt{GeminiCap-aug}) not only achieves the best performance on all general tasks (T2A, V2A, TV2A) but is also essential for enabling fine-grained control.
}
    \renewcommand{\arraystretch}{0.85}
    \resizebox{\textwidth}{!}{
    \begin{tabular}{lcccccccccccc}
    \toprule
    \multirow{2}{*}{Caption Method} & \multicolumn{3}{c}{Instruction-following T2A} & \multicolumn{2}{c}{T2A} & \multicolumn{2}{c}{V2A} & \multicolumn{2}{c}{TV2A} \\
    \cmidrule(lr){2-4} \cmidrule(lr){5-6} \cmidrule(lr){7-8} \cmidrule(lr){9-10}
    & Cat-acc $\uparrow$ & Cnt-acc $\uparrow$ & Ord-acc $\uparrow$ & IS $\uparrow$ & FAD $\downarrow$ & IS $\uparrow$ & FAD $\downarrow$ & IS $\uparrow$ & FAD $\downarrow$ \\
    \midrule
    Labels & 17.35 & 2.80 & 4.60 & 7.59 & 6.02 & 10.46 & 1.81 & 10.62 & 3.41 \\    
    AudioSetCaps & 27.85 & 6.40 & 4.80 & 10.08 & 3.19 & 11.35 & 1.33 & 12.39 & \underline{1.56} \\        
    QwenCap & 24.60 & 6.40 & 6.20 & 9.74 & 4.40 & 10.57 & 1.67 & 11.79 & 1.95 \\
    GeminiCap & \underline{28.05} & \underline{9.60} & \underline{7.60} & \underline{10.81} & \underline{3.02} & \underline{11.48} & \underline{1.31} & \underline{12.78} & 1.70 \\
    GeminiCap-aug & \textbf{28.91} & \textbf{10.20} & \textbf{7.80} & \textbf{10.93} & \textbf{2.91} & \textbf{11.69} & \textbf{1.15} & \textbf{12.90} & \textbf{1.48} \\
    \bottomrule
    \end{tabular}
    }

    \label{tab:ablation_data_aug}
\end{table*}

%% file: table/ablation_maf.tex
\begin{table*}[h]
    \centering
    \renewcommand{\arraystretch}{1}
    \small
    \caption{
    \textbf{Ablation study of the MAF architecture components.}
    We evaluate the contribution of the Gate and Query mechanisms by removing them individually.
    The results show that the \texttt{Full MAF}, which includes both components, achieves the best performance across most metrics.
    This confirms that our complete design is essential for effective multimodal fusion.
}
    \begin{tabular}{l@{\hspace{1em}}c@{\hspace{0.8em}}c@{\hspace{0.8em}}c@{\hspace{0.8em}}c@{\hspace{0.8em}}c@{\hspace{0.8em}}c@{\hspace{0.8em}}c@{\hspace{0.8em}}c@{\hspace{0.8em}}c}
    \toprule
    Components & Gate & Query & KL $\downarrow$ & IS $\uparrow$ & FD $\downarrow$ & FAD $\downarrow$ & Duration $\downarrow$ & Frequency $\downarrow$ & Ordering $\downarrow$ \\
    \midrule
    w/o MAF & $\times$ & $\times$ & 1.83 & 10.70 & 11.60 & 2.67 & 3.022 & 1.359 & 0.912 \\    
    w/o Gate & $\times$ & $\checkmark$ & \underline{1.69} & 11.66 & 9.72 & \underline{2.00} & 2.945 & 1.348 & \textbf{0.876} \\
    w/o Query & $\checkmark$ & $\times$ & 1.71 & \underline{11.72} & \underline{9.65} & 2.08 & \underline{2.841} & \underline{1.328} & 0.912 \\
    Full MAF & $\checkmark$ & $\checkmark$ & \textbf{1.68} & \textbf{11.84} & \textbf{9.64} & \textbf{1.98} & \textbf{2.827} & \textbf{1.302} & \underline{0.888} \\
    \bottomrule
    \end{tabular}

    \label{tab:maf_ablation}
\end{table*}

%% file: table/ablation_acceleration.tex
\begin{table*}[t]
    \centering
    \small
    \caption{
        \textbf{Ablation studies of efficient few-step distillation.}
        Results are reported on AudioCaps and MusicCaps.
        Best and second-best metric values within each ablation group are shown in \textbf{bold} and \underline{underlined}, respectively; bold setting names indicate the adopted configurations.
    }
    \label{tab:acceleration_ablation}
    \setlength{\tabcolsep}{5pt}
    \renewcommand{\arraystretch}{1}
    \adjustbox{max width=\textwidth}{
    \begin{tabular}{llcccccccc}
        \toprule
        \multirow{2}{*}{Ablation} & \multirow{2}{*}{Setting}
        & \multicolumn{4}{c}{AudioCaps} & \multicolumn{4}{c}{MusicCaps} \\
        \cmidrule(lr){3-6} \cmidrule(lr){7-10}
        & & KL $\downarrow$ & IS $\uparrow$ & FD $\downarrow$ & FAD $\downarrow$
        & KL $\downarrow$ & IS $\uparrow$ & FD $\downarrow$ & FAD $\downarrow$ \\
        \midrule
        \multicolumn{10}{l}{\textit{Discriminator backbone depth}} \\
        Depth & 24 blocks & 1.36 & 11.50 & \textbf{12.73} & 2.34 & \textbf{1.31} & 3.42 & 12.73 & 2.04 \\
        Depth & 18 blocks & \underline{1.35} & \textbf{12.95} & 13.95 & \textbf{1.78} & 1.35 & \textbf{3.64} & 11.73 & 2.09 \\
        Depth & 12 blocks & \underline{1.35} & 11.82 & 13.85 & \underline{2.11} & \textbf{1.31} & 3.42 & \underline{10.49} & \underline{1.75} \\
        Depth & \textbf{6 blocks} & \textbf{1.33} & \underline{12.37} & \underline{12.79} & \textbf{1.78} & \textbf{1.31} & \underline{3.61} & \textbf{9.50} & \textbf{1.54} \\
        \midrule
        \multicolumn{10}{l}{\textit{Timestep sampling probabilities}} \\
        Sampling & $(0.10,0.20,0.30,0.40)$ & 1.34 & 12.02 & 12.60 & \underline{1.80} & \underline{1.31} & 3.43 & 11.35 & 1.88 \\
        Sampling & $(0.40,0.30,0.20,0.10)$ & \underline{1.31} & 12.07 & \textbf{12.36} & 2.41 & 1.34 & \textbf{3.77} & 11.25 & 1.94 \\
        Sampling & $(0.20,0.20,0.20,0.40)$ & \textbf{1.30} & \underline{12.13} & \underline{12.44} & 1.95 & 1.33 & \underline{3.62} & 11.40 & 1.89 \\
        Sampling & $(0.40,0.20,0.20,0.20)$ & 1.37 & 12.09 & 12.97 & 1.84 & \textbf{1.30} & 3.23 & \underline{10.62} & \underline{1.87} \\
        Sampling & $\mathbf{(0.25,0.25,0.25,0.25)}$ & 1.33 & \textbf{12.37} & 12.79 & \textbf{1.78} & \underline{1.31} & 3.61 & \textbf{9.50} & \textbf{1.54} \\
        \midrule
        \multicolumn{10}{l}{\textit{Adversarial objective}} \\
        Objective & w/o GAN & \underline{1.37} & \textbf{13.19} & \underline{16.07} & \underline{2.73} & \underline{1.40} & \textbf{3.80} & \underline{15.64} & \underline{2.27} \\
        Objective & \textbf{w/ GAN} & \textbf{1.33} & \underline{12.37} & \textbf{12.79} & \textbf{1.78} & \textbf{1.31} & \underline{3.61} & \textbf{9.50} & \textbf{1.54} \\
        \bottomrule
    \end{tabular}}
\end{table*}

%% file: table/flow_vs_diffusion.tex
\begin{table}[t]
    \centering
    \small
    \caption{
        \textbf{Ablation of the training objective: flow matching v.s. diffusion.}
        Both objectives are trained under an identical backbone and budget, and evaluated on AudioCaps and MusicCaps.
        Flow matching attains quality comparable to diffusion while being naturally compatible with our few-step distillation.
        Best values are shown in \textbf{bold}.
    }
    \label{tab:flow_vs_diffusion}
    \setlength{\tabcolsep}{4pt}
    \renewcommand{\arraystretch}{1}
    \adjustbox{max width=\columnwidth}{
    \begin{tabular}{lcccccccc}
        \toprule
        \multirow{2}{*}{Objective}
        & \multicolumn{4}{c}{AudioCaps} & \multicolumn{4}{c}{MusicCaps} \\
        \cmidrule(lr){2-5} \cmidrule(lr){6-9}
        & KL $\downarrow$ & IS $\uparrow$ & FD $\downarrow$ & FAD $\downarrow$
        & KL $\downarrow$ & IS $\uparrow$ & FD $\downarrow$ & FAD $\downarrow$ \\
        \midrule
        Diffusion        & 2.19 & 10.77 & 13.47 & \textbf{1.97} & \textbf{1.36} & 3.20 & \textbf{13.46} & \textbf{1.59} \\
        \textbf{Flow matching} & \textbf{2.18} & \textbf{10.82} & \textbf{13.23} & 2.19 & 1.39 & \textbf{3.31} & 13.49 & 1.62 \\
        \bottomrule
    \end{tabular}}
\end{table}

%% file: table/music_data_ablation.tex
\begin{table}[t]
    \centering
    \small
    \caption{
        \textbf{Ablation of the music training data.}
        We compare training on the original 360K video-music subset against the full V2M-500K, evaluated on MusicCaps.
        Scaling the corpus to 500K consistently improves music generation.
    }
    \label{tab:music_data_ablation}
    \setlength{\tabcolsep}{6pt}
    \renewcommand{\arraystretch}{1}
    \adjustbox{max width=\columnwidth}{
    \begin{tabular}{lcccc}
        \toprule
        Training data & KL $\downarrow$ & IS $\uparrow$ & FD $\downarrow$ & FAD $\downarrow$ \\
        \midrule
        V2M-360K          & 1.51 & 3.21 & 12.64 & 2.15 \\
        V2M-500K & 1.45 & 3.34 & 11.57 & 1.92 \\
        \bottomrule
    \end{tabular}}
\end{table}

%% file: sec/7_Conclusion.tex
\section{Conclusion and Future Work}

\noindent\textbf{Conclusion.}
In this work, we presented \model, a unified and efficient framework for anything-to-audio generation under flexible multimodal conditions. By combining a multimodal diffusion Transformer with the Multimodal Adaptive Fusion module, \model\ supports diverse audio and music generation tasks from text, video, and audio inputs within a single model. For scalable unified training, we constructed \dataset\, a large-scale dataset with fine-grained multimodal annotations, and introduced \benchmark\ for evaluating instruction-following ability in text-to-audio generation. We further improved inference efficiency by distilling a multi-step teacher into a few-step student via Distribution Matching Distillation and a diffusion-based discriminator. Extensive experiments show that \model\ achieves strong generation quality, cross-modal alignment, and instruction following while substantially reducing sampling cost.

\noindent\textbf{Limitations and Future Work.}
Despite the strong results, \model\ has several limitations. First, both \modelbase\ and \model\ are trained on short ($10$-second) clips, restricting their applicability to long-form scenarios such as full-length film scoring or extended musical compositions. Second, the output domain is confined to general audio and music; speech, with its rich linguistic and prosodic structure, is not yet covered by our unified framework. Third, although \model\ exhibits strong fine-grained controllability, its accuracy still degrades under extreme instruction-following regimes, such as many concurrent or rapidly alternating sound events and tight timestamp tolerances. Promising directions include long-context modeling for minute- or song-level generation, unifying speech within the anything-to-audio framework, denser temporal supervision for instruction following, and flexible-step generation that adapts the number of denoising steps to the input.

%% file: sec/appendix.tex
\clearpage
\appendix
\setcounter{table}{0}   
\setcounter{figure}{0}
\renewcommand{\thetable}{A\arabic{table}}
\renewcommand{\thefigure}{A\arabic{figure}}
\section{Appendix}

\label{sec:Appendix}
\counterwithin{table}{section}
\counterwithin{figure}{section}

\input{table/dataset}

\input{table/dataset_2}

\subsection*{Appendix overview}
This appendix supplements the main paper with expanded details on our datasets, evaluation methodologies, and a broader range of experimental results. Section~\ref{sec:appendix_datasets_total} details our data sources, the two-stage annotation process, and statistics of the V2M-500K corpus. Section~\ref{sec:appendix_metrics} describes the evaluation metrics, including the quality, alignment, instruction-following, and efficiency measures used throughout the paper. Section~\ref{sec:appendix_benchmark} introduces \benchmark, our benchmark for instruction-following text-to-audio generation, together with its automated evaluation pipeline. We then present an expanded set of results in Section~\ref{sec:appendix_more_results}, including additional quantitative comparisons such as audio-visual alignment on video-to-audio.

\subsection{Datasets}
\label{sec:appendix_datasets_total}
\subsubsection{Training and test datasets}
\label{sec:appendix_datasets}
Table~\ref{tab:dataset_overview} provides an overview of all datasets used in this work. Table~\ref{tab:dataset} outlines the new captions we annotated for training and testing our unified model. We will open-source these caption datasets to facilitate further research.

\subsubsection{Further details on the \dataset\ dataset}
\label{sec:appendix_if-caps}
As described in the main text, the \dataset\ dataset is generated via a multi-step pipeline designed to produce rich, structured annotations for existing video-audio clips. This section provides a detailed breakdown of our annotation schema and showcases representative samples.

\noindent\textbf{Annotation Schema.}
Each sample in \dataset\ is accompanied by a comprehensive set of annotations designed to provide multi-faceted supervision for training. The key fields are as follows:
\begin{itemize}
\item \textbf{caption}: A holistic, high-level natural language description of the audio content, summarizing the main events and their context.
\item \textbf{category}: A structured dictionary that provides sound event classification and, where applicable, the discrete count of each event. For continuous or unquantifiable sounds (e.g., background noise, speech), the count is marked as null.
\item \textbf{SED (Sound Event Detection)}: A list providing fine-grained temporal localization. Each entry in the list maps a precise timestamp (e.g., "00:02-00:06") to a description of the sound event occurring within that specific time frame.
\item \textbf{time\_relation}: A field describing the temporal relationship between distinct sound events. This can specify a sequential order (e.g., "Event A, Event B") or more complex relationships like "interleave" for overlapping sounds.
\end{itemize}
This structured format allows our model to learn not just what sounds are present, but also how many, when, and in what order, which is critical for developing advanced instruction-following capabilities.

\noindent\textbf{Annotation Samples.}
Below is an example from \dataset\ that illustrates the richness and detail of our annotation schema, demonstrating a complex scene with overlapping, continuous, and countable events.
\begin{tcolorbox}[breakable, enhanced jigsaw, parbox=false, colback=white, colframe=black, boxrule=0.2mm, arc=0mm, title=Example, before skip=1.5mm, after skip=2mm, fonttitle=\small\ttfamily]
\begin{lstlisting}[basicstyle=\small\ttfamily, breaklines=true]
{
"caption": "A woman is speaking continuously, while a dog yips 
            twice in the background.", 
"category": {
            "Female speech": null, 
            "Yip": 2, 
            "Background noise": null
            }, 
"SED": [
        {"00:00-00:09": "A woman is speaking throughout the audio, accompanied by faint background noise."}, 
        {"00:00-00:01": "A dog lets out a yip in the background."}, 
        {"00:08-00:09": "A dog yips again in the background."}
        ], 
"time_relation": "interleave", 
"audio_id": "TATdZPmzMU8_90000"
}
\end{lstlisting}
\end{tcolorbox}

\noindent\textbf{Data Augmentation Process.}
As mentioned in the main text, a key step in our pipeline is to leverage a cost-effective model (Qwen2-Audio) to augment the initial, high-quality annotations generated by Gemini 2.5 Pro. The goal is to increase the linguistic and structural diversity of our dataset. By generating multiple, semantically equivalent but stylistically different captions for the same audio clip, we train our model to be robust to variations in user prompts and to develop a more generalized understanding of the relationship between language and sound.
The augmentation process is guided by the structured fields of the original annotation. The model is prompted to generate new captions from different perspectives: rephrasing the original description, or generating new descriptions based purely on the category and count, the SED timestamps, or the time\_relation fields.
Below, we use the second example from the previous section to illustrate this structured augmentation process.
\begin{tcolorbox}[breakable, enhanced jigsaw, parbox=false, colback=white, colframe=black, boxrule=0.2mm, arc=0mm, title=Original Audio Annotation (Generated by Gemini 2.5 Pro), before skip=1.5mm, after skip=2mm, fonttitle=\small\ttfamily]
\begin{lstlisting}[basicstyle=\small\ttfamily, breaklines=true]
{"caption": "The audio features the mechanical sound of a firearm being handled, immediately followed by two separate bursts of machine gun fire.", "category": {"Machine gun": 2, "Generic impact sounds": 1}, "SED": [...], "time_relation": "Generic impact sounds, Machine gun", "audio_id": "c9OnubhhvZY_0"}
\end{lstlisting}
\end{tcolorbox}
This single, rich annotation serves as the seed for generating a variety of new training captions, each emphasizing a different aspect of the audio content.

\begin{tcolorbox}[breakable, enhanced jigsaw, parbox=false, colback=white, colframe=black, boxrule=0.2mm, arc=0mm, title=Augmented Audio Captions (Generated by Qwen2-Audio), before skip=1.5mm, after skip=2mm, fonttitle=\small\ttfamily]
    
    \textbf{1. Caption Rephrasing (Linguistic Diversity)}
    \begin{quote}
        \textit{``A gun is cocked, followed by two bursts of machine gun fire."} \\
        \textit{``After the sharp, metallic sound of a firearm mechanism, two rapid-fire bursts from a machine gun are heard."}
    \end{quote}

    \textbf{2. Augmentation from Category and Count}
    \begin{quote}
        \textit{``The audio contains two sounds of a machine gun and one generic impact sound."}
    \end{quote}

    \textbf{3. Augmentation from SED}
    \begin{quote}
        \textit{``The sound of a firearm being handled is audible for the first second, followed by a burst of machine gun fire from 1 to 5 seconds and a second burst from 6 to 8 seconds."}
    \end{quote}

    \textbf{4. Augmentation from Time Relation}
    \begin{quote}
        \textit{``In this audio, the sound of a generic impact occurs first, followed by two distinct machine gun sounds."}
    \end{quote}

\end{tcolorbox}
This structured augmentation strategy ensures our model is exposed to a wide variety of textual descriptions, learning to associate not only high-level captions but also explicit instructions about count, timing, and order with the corresponding audio features. Similarly, for music data, this process generates varied descriptions of genre, mood, instrumentation, and tempo, teaching the model to comprehend both high-level artistic direction and specific musical components.

\begin{tcolorbox}[breakable, enhanced jigsaw, parbox=false, colback=white, colframe=black, boxrule=0.2mm, arc=0mm, title=Original Music Annotation (Generated by Gemini 2.5 Pro), before skip=1.5mm, after skip=2mm, fonttitle=\small\ttfamily]
\begin{lstlisting}[basicstyle=\small\ttfamily, breaklines=true]
{"caption": "A heartwarming acoustic track featuring a blend of softly strummed guitar and a simple, melodic piano line, creating a gentle and uplifting atmosphere.", "genre": "Acoustic Pop, Instrumental", "mood": "Heartwarming, Gentle, Uplifting", "instrument": ["Acoustic Guitar", "Piano"], "tempo": "Slow to Moderate"}
\end{lstlisting}
\end{tcolorbox}

This structured music annotation is then used to generate diverse new training captions, each focusing on a different attribute:
\begin{tcolorbox}[breakable, enhanced jigsaw, parbox=false, colback=white, colframe=black, boxrule=0.2mm, arc=0mm, title=Augmented Music Captions (Generated by Qwen2-Audio), before skip=1.5mm, after skip=2mm, fonttitle=\small\ttfamily]

\textbf{1. Caption Rephrasing}
\begin{quote}
    \textit{``A gentle instrumental piece with the interwoven sounds of an acoustic guitar and piano."} \\
    \textit{``Soft guitar strumming and a simple piano melody combine to create an uplifting acoustic pop track."}
\end{quote}

\textbf{2. Augmentation from Genre}
\begin{quote}
    \textit{``An instrumental acoustic pop track featuring piano and guitar."}
\end{quote}

\textbf{3. Augmentation from Mood}
\begin{quote}
    \textit{``A heartwarming, gentle, and uplifting piece of music featuring acoustic guitar and piano."}
\end{quote}

\textbf{4. Augmentation from Tempo}
\begin{quote}
    \textit{``A slow to moderate tempo instrumental track with piano and acoustic guitar."}
\end{quote}
\end{tcolorbox}

\subsubsection{Statistics of the V2M-500K dataset}
\label{sec:appendix_v2m500k}
\begin{figure*}[t]
    \centering
    \includegraphics[width=0.95\linewidth]{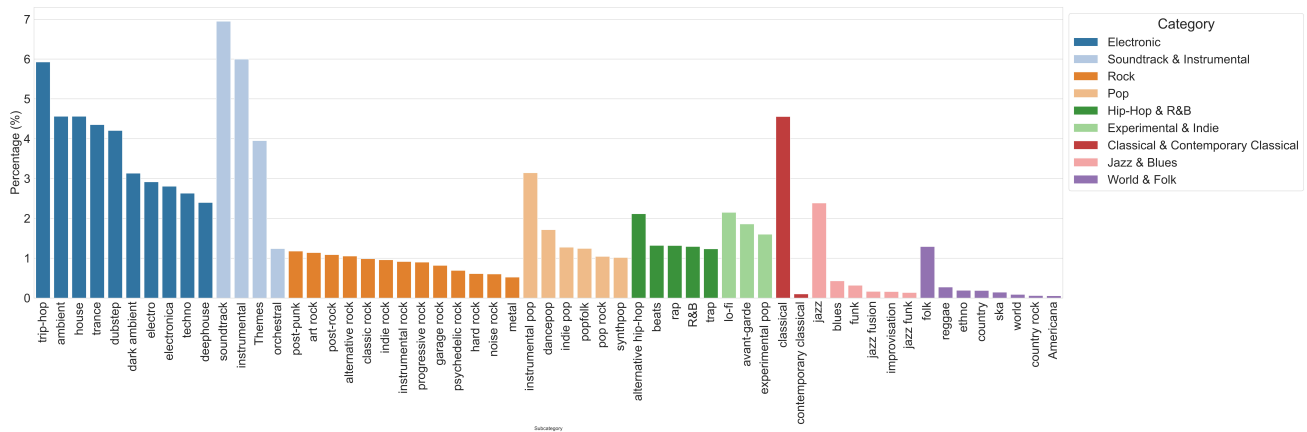}
    \caption{
    Distribution of music genres in V2M-500K, showcasing the diverse representation of genres such as electronic, classical, and jazz.
    }
    \label{fig:v2m_categories}
\end{figure*}

\begin{figure*}[t]
    \centering
    \includegraphics[width=0.95\linewidth]{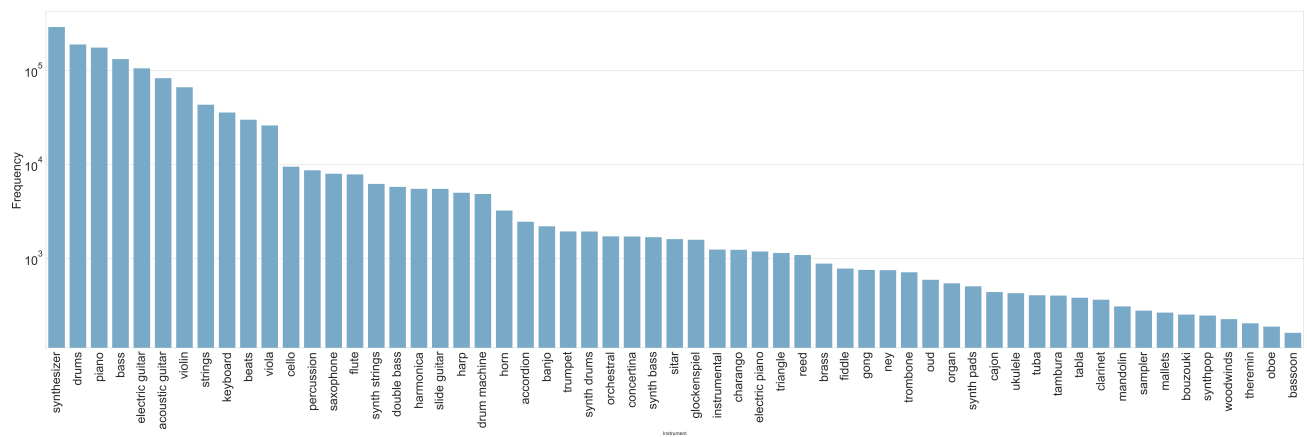}
    \caption{Distribution of instruments in V2M-500K, emphasizing the frequent usage of synthesizers, pianos, and drums, while also including diverse instruments such as violins and saxophones.}
    \label{fig:v2m_instruments}
\end{figure*}

To provide a closer look at the composition of our \textbf{V2M-500K} corpus, we visualize the distribution of musical genres and instruments in Figures~\ref{fig:v2m_categories} and~\ref{fig:v2m_instruments}. The genre and instrument tags are obtained from the structured fields produced by our two-step annotation pipeline (Sec.~\ref{sec:appendix_if-caps}); a single clip can be associated with multiple tags, so the counts reflect occurrences rather than disjoint partitions of clips.

\noindent\textbf{Genre and subgenre coverage.} Figure~\ref{fig:v2m_categories} groups fine-grained subgenres into nine high-level categories. V2M-500K spans a broad spectrum of musical styles, with \emph{Soundtrack \& Instrumental}, \emph{Electronic} and \emph{Classical} forming the head of the distribution, while sizeable presences of \emph{Pop}, \emph{Rock}, \emph{Hip-Hop \& R\&B}, \emph{Experimental \& Indie}, \emph{World \& Folk}, and \emph{Jazz \& Blues} together provide a long-tailed coverage that is essential for training a generalist video-to-music model. Compared with the original V2M-360K corpus introduced in VidMuse~\citep{tian2025vidmuse}, V2M-500K substantially enlarges every category while preserving the overall genre balance.

\noindent\textbf{Instrumentation.} Figure~\ref{fig:v2m_instruments} reports the per-instrument frequency on a logarithmic scale (instruments occurring fewer than 140 times are omitted for clarity). The distribution is dominated by common popular-music instruments such as \emph{synthesizer}, \emph{piano}, \emph{drums}, \emph{bass} and \emph{electric/acoustic guitar}, and gradually transitions into orchestral and folk instruments (e.g., \emph{strings}, \emph{flute}, \emph{cello}, \emph{saxophone}) and finally a long tail of region-specific instruments (e.g., \emph{sitar}, \emph{oud}, \emph{tabla}, \emph{bouzouki}, \emph{qanun}). This long-tailed yet wide-coverage instrumentation enables \model\ to follow fine-grained instrument cues and generalize to under-represented musical contexts.

\subsection{Details of evaluation metrics}
\label{sec:appendix_metrics}

\begin{figure*}[!t]
  \centering
  \includegraphics[width=0.95\linewidth]{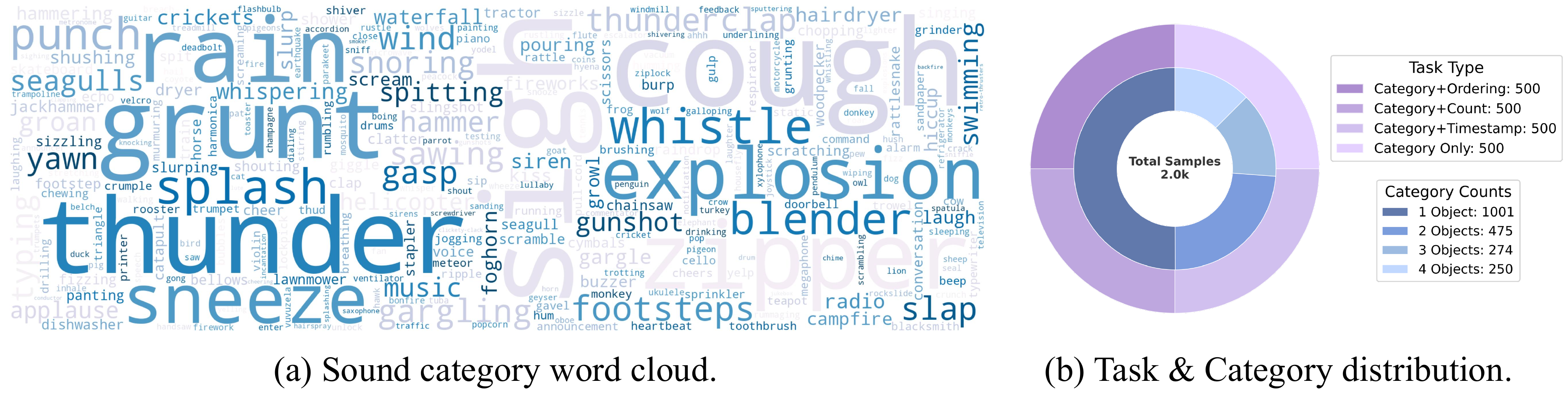}   
  \caption{The composition of the \benchmark\ benchmark. (a) Word cloud of sound event categories. (b) Distribution of task types and category counts.}
  \label{fig:benchmark}   
\end{figure*}

\noindent\textbf{Fréchet Audio Distance (FAD).} To evaluate the perceptual quality of the generated audio, we employ FAD, a reference-free metric analogous to the FID \citep{heusel2017gans} score used in image generation. The metric functions by comparing the statistical distance between embedding distributions of generated audio and real-world audio. A smaller distance suggests the generated audio is of higher acoustic quality. For our calculations, we utilize the VGGish \citep{hershey2017cnn} feature extractor.

\noindent\textbf{Fréchet Distance (FD).} While similar in principle to FAD, FD serves as a distinct measure of audio similarity by employing a different feature extractor. We use an FD variant based on PANNs \citep{kong2020panns} embeddings. Given that PANNs models are pretrained on the extensive AudioSet \citep{gemmeke2017audio}, this metric is considered to be highly robust for evaluating audio fidelity.

\noindent\textbf{Kullback-Leibler Divergence (KL).} The KL divergence is used to approximate the acoustic similarity between generated and reference audio samples. This is achieved by measuring the divergence between the multi-label class prediction distributions produced by a PANNs model for both sets of samples.

\noindent\textbf{Inception Score (IS).}
The IS is a widely used metric to evaluate the performance of generative models. Besides assessing the diversity of the generated samples, IS also evaluates their quality, measuring the clarity and recognizability of individual audio events \citep{donahue2018adversarial, majumder2024tango, liu2023audioldm}. Given its ability to provide a single, holistic score reflecting both of these aspects without needing a reference prompt, we selected IS as the unified metric for the comprehensive performance comparison in our teaser Fig.~\ref{fig:teaser}. This allows for a fair and consistent visualization of our model's capabilities across the wide array of supported tasks.

\noindent\textbf{ImageBind Score \citep{girdhar2023imagebind}.} We assess the semantic alignment between generated audio and conditioning videos using the ImageBind Score. This score is calculated as the cosine similarity between the audio and video embeddings from the respective branches of the ImageBind model. 

\noindent\textbf{CLAP Score.} The Contrastive Language-Audio Pretraining (CLAP) model \citep{elizalde2023clap} learns a joint embedding space where audio clips and their corresponding text descriptions are aligned. We use it to evaluate the semantic alignment between generated audio and a text prompt, calculated as the cosine similarity between their embeddings from the pretrained CLAP encoders \citep{wu2023large}. A higher score indicates better alignment.

\noindent\textbf{Production Complexity (PC) and Production Quality (PQ).} These metrics are derived from the Meta Audiobox Aesthetics framework \citep{tjandra2025meta}. 
PQ focuses on the technical aspects of an audio recording, such as its clarity, fidelity, dynamics, and frequency balance. In contrast, PC evaluates the complexity of an audio scene by measuring the number of distinct audio components present, such as multiple instruments or the co-occurrence of speech, music, and sound effects. Both are no-reference metrics, allowing the assessment of individual audio clips without needing a ground-truth comparison sample.

\noindent\textbf{Ordering, Duration, Frequency, and Timestamp.} These metrics are components of the STEAM evaluation framework, proposed in the AudioTime \citep{xie2025audiotime} to assess the temporal controllability of audio generation models. Ordering is an error rate that measures whether sound events are generated in the specified sequence. Duration and Frequency are calculated as the L1 error between the specified and detected event durations and occurrence counts, respectively. Timestamp evaluates the precise timing of events (onset and offset) using the F1-score, a common metric in sound event detection.

\noindent\textbf{Category, Count, Ordering, and Timestamp accuracy.} See ~\ref{sec:appendix_benchmark}.

\noindent\textbf{Alignment accuracy (AlignAcc) and audio-visual synchronization (AVSync).} For video-to-audio generation, we further assess audio-visual synchronization from two complementary angles. AlignAcc \citep{luo2024diff} reports the fraction of generated samples judged as a real audio-visual pair by a pretrained alignment classifier (higher is better); the classifier is trained against both temporally shifted and mismatched pairs, so it captures semantic relevance and temporal synchronization jointly. AVSync \citep{cheng2025mmaudio} instead measures the temporal offset estimated by Synchformer \citep{iashin2024synchformer}, where a value closer to zero indicates tighter synchronization.

\noindent\textbf{Efficiency (NFE, Latency, and RTF).} We report three complementary efficiency measures. The number of function evaluations (NFE) counts the forward passes per sample as a hardware-independent compute proxy; since classifier-free guidance doubles the passes per step, CFG-based methods (all baselines and \modelbase) have $\text{NFE}=2\times\text{Steps}$, whereas the CFG-free \model\ has $\text{NFE}=\text{Steps}$. Latency is the end-to-end wall-clock time per clip, measured on a single NVIDIA 4090 GPU at batch size 1 (mean$\pm$std over 20 runs after 5 warm-ups). The real-time factor (RTF) is latency divided by audio duration, where lower is better and RTF $<\!1$ means faster than real time.

\noindent\textbf{Overall Quality (OVL) and Relevance (REL).} For our subjective evaluation, 10 professional audio experts rated each generated sample on a scale of 1 to 100 on two standard criteria. OVL assesses the intrinsic perceptual fidelity of the audio itself—focusing on aspects like clarity and freedom from artifacts—independent of the prompt. In parallel, REL measures the semantic alignment between the audio and its conditioning input, evaluating how accurately the content reflects the instructions from the provided text or video. This evaluation protocol follows the established methodologies of prior work \citep{kreuk2022audiogen, liu2023audioldm}. Example of the questionnaire interface is shown in Table~\ref{tab:user_study_questionnaire}.

\begin{table*}[h]
\centering
\caption{Simplified example of the questionnaire for human evaluation, showcasing the four main task types. Experts provided scores for OVL and REL.}
\begin{tabular}{l l c c}
\toprule
\textbf{File Name} & \textbf{Prompt (Text or Video)} & \textbf{OVL (1-100)} & \textbf{REL (1-100)} \\
\midrule
9964.wav & A loud white noise and then some beeping. & 55 & 65 \\
0928.wav & An uplifting folk-pop instrumental track. & 70 & 55 \\
1441.wav & {[Video of a person walking on dry leaves]} & 80 & 70 \\
1701.wav & {[Video of a drone shot over a sunrise mountain]} & 65 & 60 \\
... & ... & ... & ... \\
\bottomrule
\end{tabular}

\label{tab:user_study_questionnaire}
\end{table*}

\subsection{Benchmark and metrics for instruction-following in T2A}
\label{sec:appendix_benchmark}
To rigorously and scalably evaluate the instruction-following capabilities of Text-to-Audio generation models, we introduce a new benchmark, \textbf{\benchmark}, and a corresponding automated evaluation pipeline. This framework is designed to dissect a model's ability to adhere to complex compositional instructions.

\noindent\textbf{\benchmark\ Composition and Design.}
\benchmark\ is a prompt-based benchmark comprising 2k challenging, natural language prompts generated by Gemini 2.5 Pro. It is structured to systematically probe four key dimensions of controllability. As illustrated in Figure~\ref{fig:benchmark}, our benchmark encompasses a diverse vocabulary of sound categories and a balanced task structure to enable a rigorous and comprehensive evaluation. The benchmark is divided into four task types, each containing 500 prompts:
\begin{itemize}
\item \textbf{Category-only}: Evaluates the generation of correct sound events. Prompts contain between one and five distinct sound categories (100 prompts for each count).
\item \textbf{Category+Count}: Assesses the ability to generate a precise number of sound events. To avoid ambiguity, prompts in this category feature only a single sound type, with the required count ranging from one to five (100 prompts for each count).
\item \textbf{Category+Ordering}: Measures adherence to temporal sequence. Prompts specify an order for either two or three distinct sound categories.
\item \textbf{Category+Timestamp}: Tests temporal localization. To ensure clarity, prompts specify a start and end time for a single sound category.
\end{itemize}
Below are representative examples for each task type, including the prompt and its corresponding structured metadata.
\begin{tcolorbox}[breakable, enhanced jigsaw, parbox=false, colback=white, colframe=black, boxrule=0.2mm, arc=0mm, title=\benchmark\ Examples, before skip=1.5mm, after skip=2mm, fonttitle=\small\ttfamily]
\begin{lstlisting}[basicstyle=\small\ttfamily, breaklines=true]
{
"id": "T2A_01565", 
"type": "category-only", 
"prompt": "A violent storm at sea, with a loud clap of thunder and a huge wave crashing over the deck.", 
"category": "thunder, wave crash"
}
{
"id": "T2A_00031", 
"type": "category+count", 
"prompt": "A single, loud bark from a dog in the distance.", 
"category": "dog bark", 
"count": {"dog bark": 1}
}
{
"id": "T2A_00575", 
"type": "category+ordering", 
"prompt": "The sound of a person gargling, followed by the splash of water in the sink.", 
"category": "gargle, water splash", 
"time_relation": "gargle, water splash"
}
{
"id": "T2A_01105", 
"type": "category+timestamp", 
"prompt": "The sound of a crowd cheering is present from 2.0 seconds to 6.0 seconds.", 
"category": "crowd cheering", 
"timestamp": {"crowd cheering": {"start": 2.0, "end": 6.0}}
}

\end{lstlisting}
\end{tcolorbox}
\noindent\textbf{Evaluation Metrics.}
Corresponding to the benchmark's structure, we define four strict, accuracy-based metrics: Category Accuracy (Cat-acc), Count Accuracy (Cnt-acc), Ordering Accuracy (Ord-acc), and Timestamp Accuracy (TS-acc). The final score for each metric is the percentage of ``correct" judgments.
\begin{itemize}
\item \textbf{Cat-acc}: A judgment is ``correct" only if \textit{all} sound categories specified in the prompt are detected in the generated audio. This is evaluated on all 2,000 samples.
\item \textbf{Cnt-acc}: A judgment is ``correct" only if the detected count for the specified category exactly matches the prompt's instruction.
\item \textbf{Ord-acc}: A judgment is ``correct" only if the detected temporal order of sound events exactly matches the specified sequence.
\item \textbf{TS-acc}: A judgment is ``correct" only if the detected event's start and end times fall within a 1-second tolerance window of the target times specified in the prompt.
\end{itemize}
\noindent\textbf{Automated Evaluation Pipeline.}
To ensure objective and scalable evaluation while preventing information leakage, we designed a novel two-step pipeline that leverages the state-of-the-art audio understanding of a powerful Multimodal Large Model (MLLM), Gemini 2.5 Pro, as an automated judge.

\begin{itemize}
\item \textbf{Step 1: Blind Audio Annotation.} In the first step, the MLLM judge receives \textit{only} the audio sample generated by the model under evaluation. It performs a blind, detailed analysis to produce a structured annotation of the audio's content. This annotation includes detected sound categories, their counts, temporal relationships, and precise sound event detection (SED) timestamps. For sounds where counting is ambiguous (e.g., continuous water flow) or ordering is not distinct, the corresponding fields are populated with null.
\begin{tcolorbox}[breakable, enhanced jigsaw, parbox=false, colback=white, colframe=black, boxrule=0.2mm, arc=0mm, title=Example of Step 1 Output (Structured Annotation), before skip=1.5mm, after skip=2mm, fonttitle=\small\ttfamily]
\begin{lstlisting}[basicstyle=\small\ttfamily, breaklines=true]
{
"caption": "The audio contains two distinct loud sounds. First, there is a deep, rolling thunderclap. After a brief pause, a powerful and sudden explosion is heard.", 
"category": {"Thunder": 1, "Explosion": 1}, 
"SED": [
        {"00:01.734-00:03.514": "A deep, rolling thunderclap is heard."}, 
        {"00:08.241-00:09.511": "A loud and sudden explosion with a distinct boom."}], 
"time_relation": "Thunder, Explosion", 
}
\end{lstlisting}
\end{tcolorbox}
\item \textbf{Step 2: LLM-based Judgment.} In the second step, the MLLM judge is provided with the original prompt from \benchmark\ and the structured annotation generated in Step 1. Acting like an examiner with an answer key, the MLLM compares the annotated audio content against the prompt's instructions. It then outputs a binary score (1 for correct, 0 for incorrect) for the relevant metric, along with a detailed textual analysis explaining its decision.
\begin{tcolorbox}[breakable, enhanced jigsaw, parbox=false, colback=white, colframe=black, boxrule=0.2mm, arc=0mm, title=Example of Step 2 Output (Final Judgment), before skip=1.5mm, after skip=2mm, fonttitle=\small\ttfamily]
\begin{lstlisting}[basicstyle=\small\ttfamily, breaklines=true]
{
"prompt": "A medieval battlefield, with the sound of a catapult launching a stone and the subsequent explosion."
"prediction": {"cat_acc": 0, "cnt_acc": null, "ord_acc": null, "ts_acc": null, 
"analysis": "The audio contains a clear and prominent sound of thunder, which is audible from the beginning and culminates in a loud clap around 00:03. However, the required category 'wave crash' is missing. While there is a sound of water starting around 00:05, it is acoustically identifiable as heavy rain rather than a distinct, powerful wave crashing."}
}
\end{lstlisting}
\end{tcolorbox}

\end{itemize}

In summary, our framework, combining \benchmark, fine-grained metrics, and a robust two-step evaluation pipeline, provides a comprehensive and replicable methodology for quantifying the instruction-following capabilities of T2A models. We will open-source our proposed benchmark and evaluation pipeline to facilitate future research in this area.

\subsection{More results}
\label{sec:appendix_more_results}
\subsubsection{Comparison results}
\label{sec:appendix_comparisons}

\noindent\textbf{Video-to-audio generation on AVVP.} Since AVVP~\citep{tian2020unified} is not seen during training, this experiment evaluates the out-of-domain generalization of \model\ under both Video-to-Audio (V2A) and Text-and-Video-to-Audio (TV2A) settings. As shown in Table~\ref{tab:avvp_v2a}, both \modelbase\ and \model\ remain highly competitive on this unseen dataset, and the few-step \model\ stays on par with the multi-step \modelbase, indicating that our framework generalizes robustly to unseen distributions and retains this robustness after distillation.

\input{table/avvp_v2a}

\noindent\textbf{Audio-visual alignment on V2A.}
Beyond standard quality metrics, video-to-audio generation places strong demands on \emph{audio-visual correspondence}, i.e., whether the generated sound is semantically faithful to and temporally synchronized with the visual content. We therefore complement the VGGSound V2A evaluation with two dedicated metrics, AlignAcc and AVSync. As shown in Table~\ref{tab:v2a_align}, \model\ attains the best AlignAcc and FAD among all methods while remaining competitive on AVSync, indicating that, despite operating in only $4$ steps, it produces video-conditioned audio that is both semantically and temporally well aligned with the input.

\input{table/v2a_align}

\noindent\textbf{Audio inpainting.} As shown in Table~\ref{tab:inpainting_results}, we conducted experiments on audio inpainting tasks, where our model outperformed the baselines \citep{liu2023audioldm, liu2024audioldm} on the AudioCaps \citep{kim2019audiocaps} and AVVP \citep{tian2020unified} test datasets. Additionally, to explore audio inpainting with various input modalities, we performed experiments on unconditioned, video-guided, and text-and-video-guided audio inpainting tasks (on AVVP). The results indicate that both text and video can effectively guide the audio inpainting task, with text providing better guidance than video. When both text and video are conditioned, the model can integrate the two modalities to achieve superior results.

\input{table/audio_inpainting}

\noindent\textbf{Music Completion.} Music completion is a task where the model generates music from a given music clip. We evaluate our model on the V2M-bench \citep{tian2025vidmuse} dataset. The results are shown in Table~\ref{tab:music_comp}. We find that our model can generate music that extends the input music clip. As the number of input modalities increases, the model's performance improves, demonstrating its strong inter-modal learning capability and ability to leverage multi-modal information to generate better music.

\input{table/music_comp}

\noindent\textbf{Image-to-audio generation.} To evaluate our model's performance on \textbf{zero-shot image-to-audio generation task}, we conducted experiments using the same settings as in \citep{xing2024seeing}. We compared our model with Seeing\&Hearing \citep{xing2024seeing} and Im2Wav \citep{sheffer2023hear}, and also constructed a baseline by combining an image caption model \citep{bai2023qwen} with a text-to-audio model \citep{majumder2024tango}. The results are shown in Table~\ref{tab:image2audio_comparison} in the Appendix. We find that our model demonstrates excellent performance in the image-to-audio generation task even without any specific training with image data.

\begin{figure*}[t]
  \centering
  \includegraphics[width=0.6\linewidth]{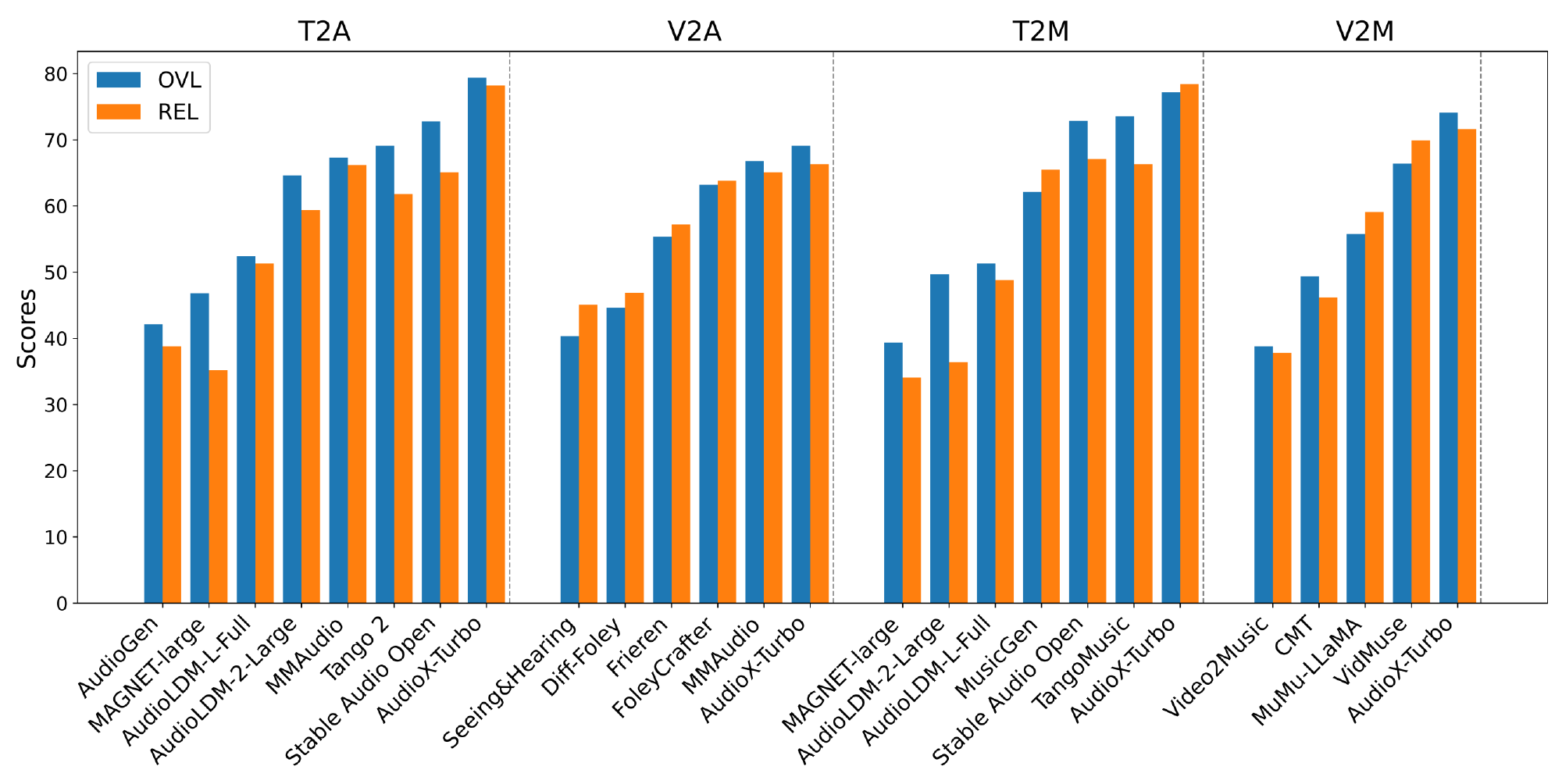}
  \caption{User study results of generated audio and music. The values represent the average OVL and REL scores across Text-to-Audio (on AudioCaps), Text-to-Music (on MusicCaps), Video-to-Audio (on VGGSound), Video-to-Music (on V2M-bench).}
  \label{fig:user_study}
\end{figure*}

\noindent\textbf{Image-to-audio generation.} To evaluate the model's capability in handling static visual inputs, we conduct a \textbf{zero-shot image-to-audio generation} experiment. Adopting the protocol of Seeing\&Hearing \citep{xing2024seeing}, we evaluate on 3k clips from the VGGSound test set, where keyframes were processed using AnimeGANv2 \citep{AnimeGANv2} to transfer them into ``Paprika style'' prior to generation. For comparison, we benchmark AudioX against Seeing\&Hearing \citep{xing2024seeing}, Im2Wav \citep{sheffer2023hear}, and a baseline combining an image caption model \citep{bai2023qwen} with a text-to-audio model \citep{majumder2024tango}. The results are shown in Table~\ref{tab:image2audio_comparison} in the Appendix. We find that our model demonstrates excellent performance in the image-to-audio generation task even without any specific training with image data.

\begin{figure*}[t]
  \centering
  \includegraphics[width=0.6\linewidth]{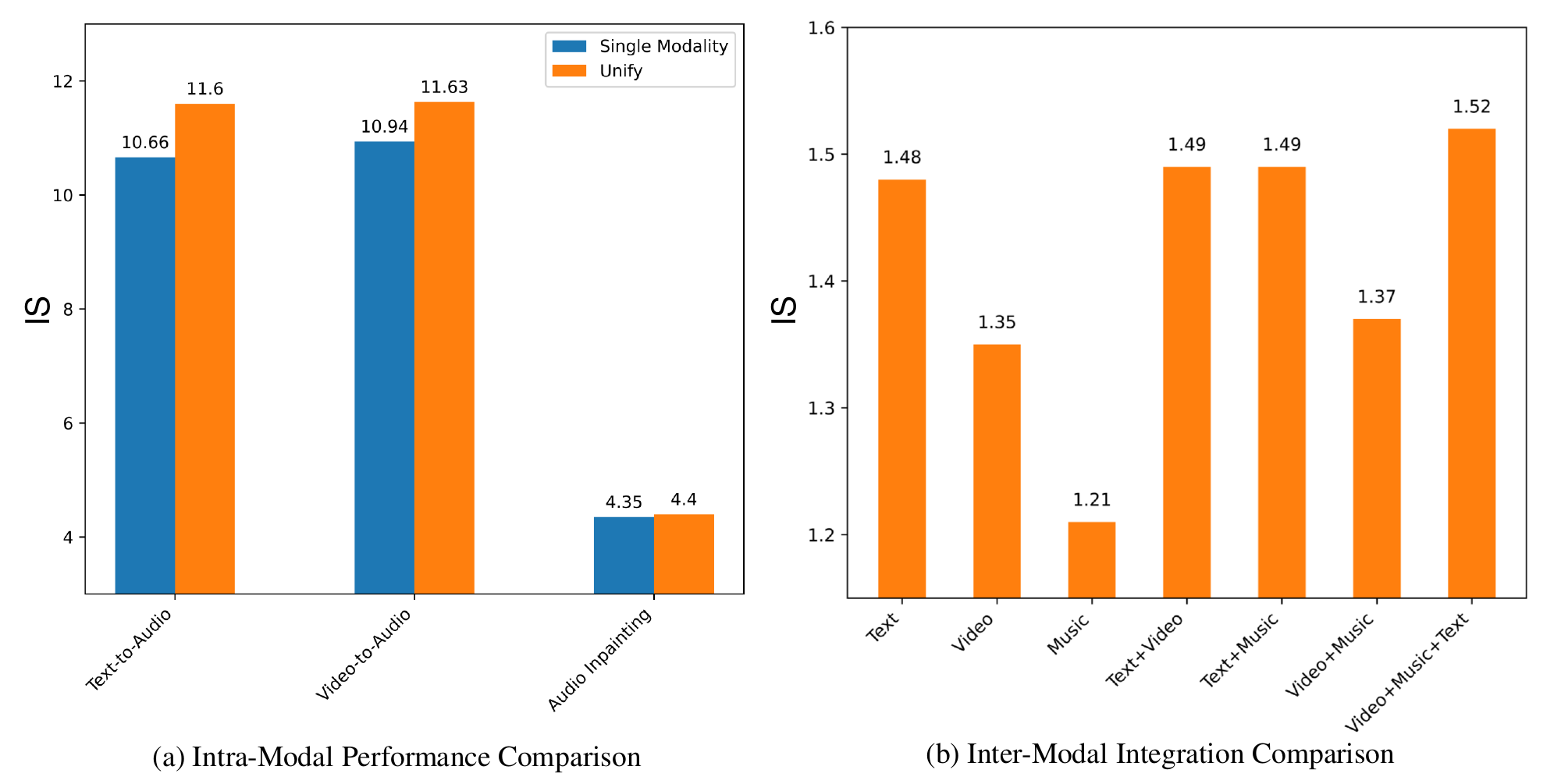}
  \caption{Ablation study comparing intra-modal and inter-modal performance of the unified model. The left compares single-modality models on text-to-audio, video-to-audio, and audio inpainting tasks. The right shows the effect of adding modalities on music generation, with performance improvements noted for each added modality. Results are based on the Inception Score (IS) metric.}
  \label{fig:ablation_modality}
\end{figure*}

\input{table/I2A}

\subsubsection{Ablation results}
\label{sec:appendix_ablation}

\noindent\textbf{Unified model performance.}

We investigate our unified model's intra- and inter-modal performance in Fig.~\ref{fig:ablation_modality}. For the intra-modal study, we compare our single unified model against specialist models trained on individual tasks (T2A, V2A, and audio inpainting). The results show our unified model consistently outperforms these specialist models, demonstrating strong intra-modal capabilities. For the inter-modal study on music generation, we find that performance progressively improves as more conditioning modalities are added (e.g., from video-only to video+text). This confirms the model's robust ability to effectively integrate multiple modalities to enhance generation quality.

\noindent\textbf{Effect of the MMDiT backbone.}
We qualitatively compare the DiT backbone used in the conference version with the MMDiT backbone adopted here. We observe that MMDiT produces audio that is better aligned with the conditioning signals, with the difference being most noticeable under video conditioning. 
Qualitative V2A examples from both backbones are provided in the supplementary material for comparison.

\subsection{Qualitative Results}
\label{sec:appendix_qualitative}
\begin{figure*}[t]
  \centering
  \begin{tabular}{cc}
    \includegraphics[width=0.42\linewidth]{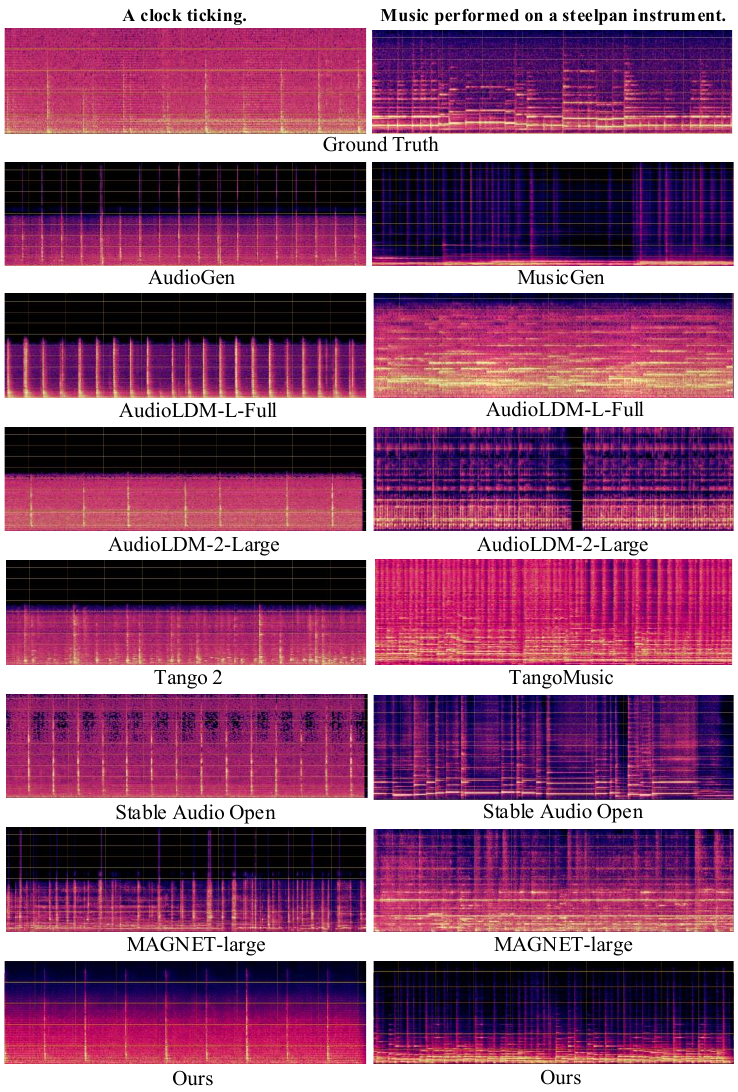} &
    \includegraphics[width=0.42\linewidth]{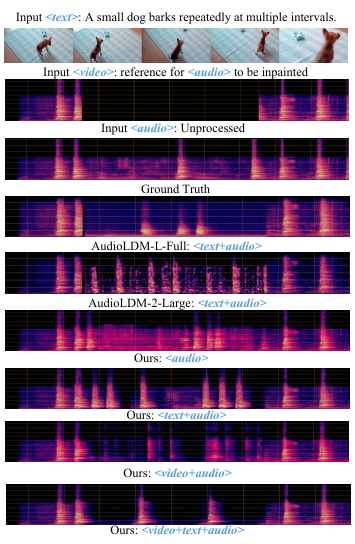} \\
    (a) T2A and T2M Results. &
    (b) Inpainting Results. \\
    \multicolumn{2}{c}{
      \includegraphics[width=0.84\linewidth]{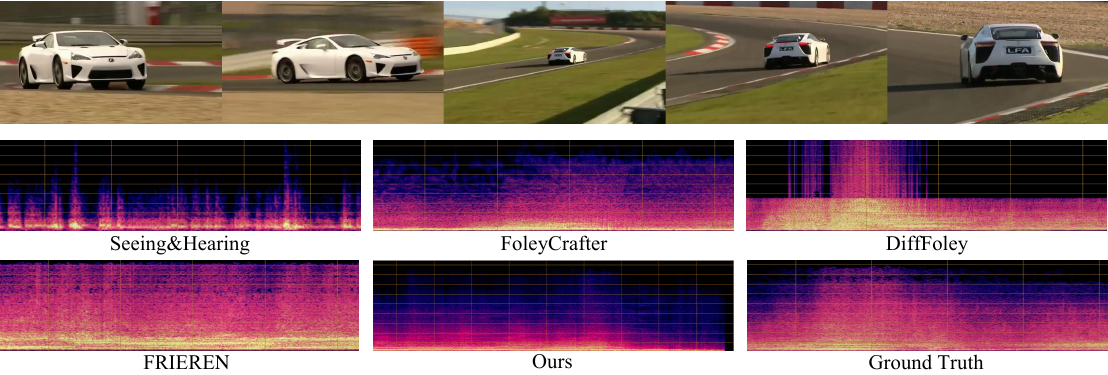}
    } \\
    \multicolumn{2}{c}{(c) V2A Results.}
  \end{tabular}
  \caption{Qualitative comparison across various tasks:  (a) In Text-to-Audio (T2A) and Text-to-Music (T2M) tasks, our model uniquely excels by consistently generating the ``ticking'' sound of a clock and accurately following the prompt "Music performed on a steelpan instrument," outperforming baselines in both rhythmic precision and genre fidelity. (b) Audio inpainting results demonstrate our model's strong context-aware capabilities and its ability to effectively integrate different input modalities. (c) Video-to-Audio (V2A) results show our model's proficiency in capturing dynamic motion sounds, such as the immersive ``drifting'' of a car, providing a richer auditory experience compared to baselines.}
\label{fig:comparison_figure}
\end{figure*}

\begin{figure*}[t]
  \centering
  \begin{tabular}{cc}
    \multicolumn{2}{c}{
      \includegraphics[width=0.84\linewidth, height=0.29\textheight]{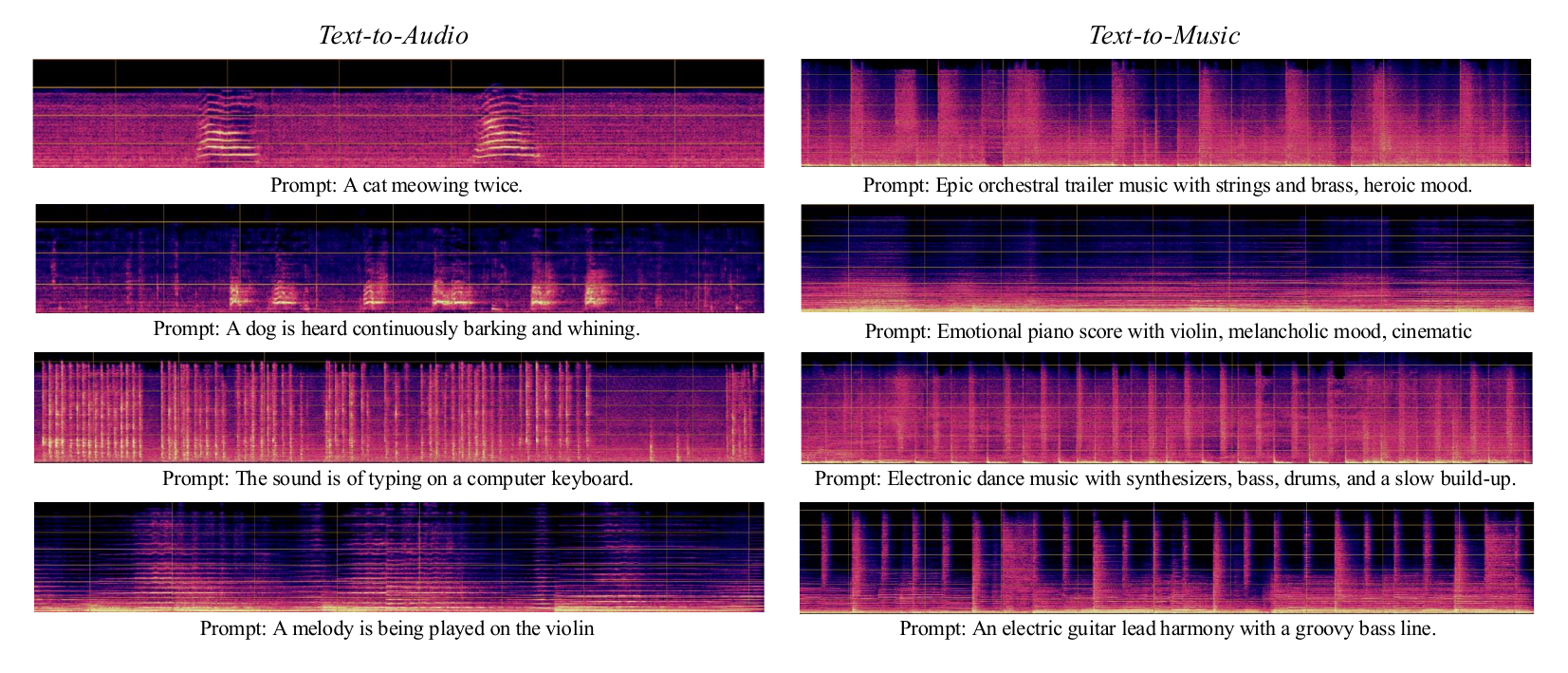}
    } \\[-1ex]
    \multicolumn{2}{c}{(a) Text-to-Audio and Text-to-Music} \\
  
    \multicolumn{2}{c}{
      \includegraphics[width=0.84\linewidth, height=0.3\textheight]{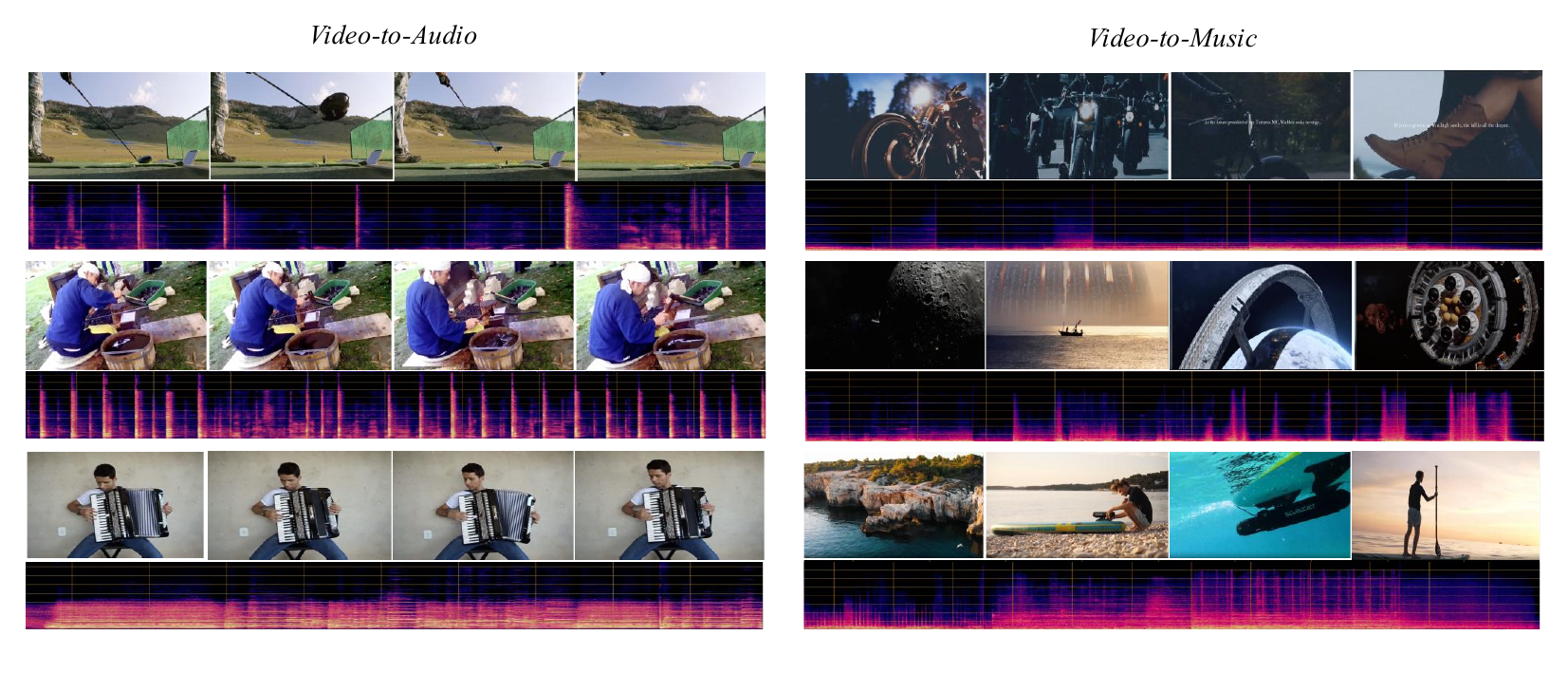}
    } \\[-1ex]
    \multicolumn{2}{c}{(b) Video-to-Audio and Video-to-Music} \\
  
    \multicolumn{2}{c}{
      \includegraphics[width=0.84\linewidth, height=0.3\textheight]{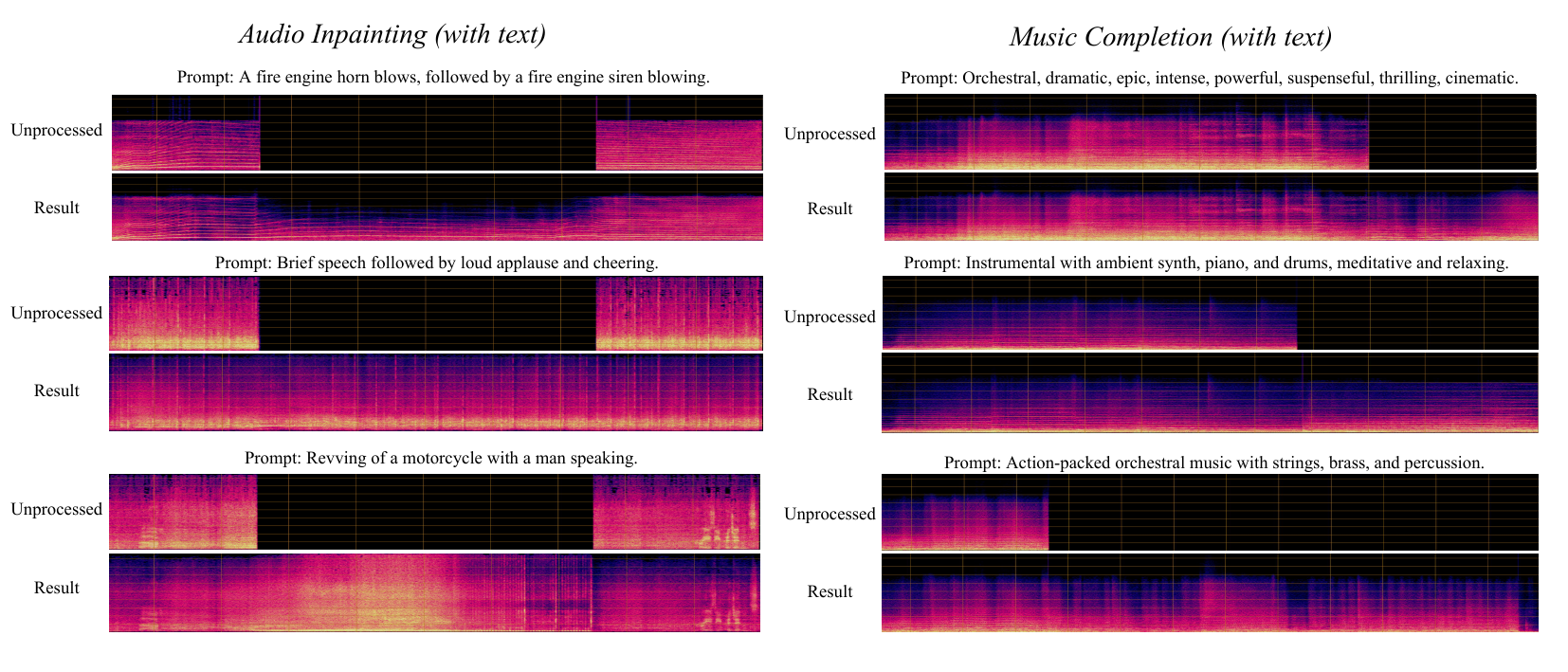}
    } \\[-1ex]
    \multicolumn{2}{c}{(c) Audio Inpainting and Music Completion}
  \end{tabular}
  \caption{Comprehensive qualitative analysis of our model's performance across various tasks: (a) Text-to-Audio and Text-to-Music synthesis, (b) Video-to-Audio and Video-to-Music generation, and (c) Audio Inpainting and Music Completion.}
\label{fig:more_result}
\end{figure*}

Figures~\ref{fig:comparison_figure} and \ref{fig:more_result} present comprehensive qualitative results.

%% file: table/dataset.tex
\begin{table*}[h]
    \centering
    \small
    \caption{Comprehensive overview of training and test datasets, detailing the number of clips (\# Clips), average duration per clip (Dur./Clip in seconds), and total duration (Dur. in hours) for each task and split. T2A: Text-to-Audio, V2A: Video-to-Audio, TV2A: Text-and-Video-to-Audio, T2M: Text-to-Music, V2M: Video-to-Music, TV2M: Text-and-Video-to-Music.}
    \renewcommand{\arraystretch}{0.95}
    \resizebox{0.8\textwidth}{!}{
    \begin{tabular}{l l l c c c}
    \toprule
    Split & Task & Dataset & \# Clips & Dur./Clip (s) & Dur. (h) \\
    \midrule
    \multirow{15}{*}{Train} 
    & \multirow{4}{*}{T2A} & AudioCaps & 45.0k & 10 & 125.1 \\
    & & WavCaps & 108.3k & 10 & 300.8 \\
    & & {\dataset} & 1268k & 10 & 3524.4 \\
    & & AudioTime & 20k & 10 & 355.5 \\    
    \cmidrule(lr){2-6}
    & \multirow{3}{*}{V2A} & VGGSound & 176.9k & 10 & 491.4 \\
    & & AudioSet Strong & 67.3k & 10 & 187.1 \\
    & & Greatest Hits & 1.0k & 10 & 2.7 \\
    \cmidrule(lr){2-6}
    & \multirow{2}{*}{TV2A} & {\dataset} & 1268k & 10 & 3524.4 \\
    & & Greatest Hits & 1.0k & 10 & 2.7 \\
    \cmidrule(lr){2-6}
    & \multirow{3}{*}{T2M} & Private & 175.2k & 240 & 11679.3 \\
    & & V2M & 7880.2k & 10 & 21889.3 \\
    & & MUCaps & 22.0k & 208 & 1273.6 \\
    \cmidrule(lr){2-6}
    & V2M & V2M & 7880.2k & 10 & 21889.3 \\
    & TV2M & V2M & 7880.2k & 10 & 21889.3 \\
    \cmidrule(lr){2-6}
    & Audio Inpainting & All audio data & 398.5k & 10 & 1107.1 \\
    \cmidrule(lr){2-6}    
    & Music Completion & All music data & 5882.9k & ~17.6 & 28746.5 \\
    \midrule
    \multirow{12}{*}{Test} 
    & \multirow{4}{*}{T2A} & AudioCaps & 4,875 & 10 & 13.5 \\
    & & VGGSound & 14,931 & 10 & 41.5 \\
    & & {\benchmark} & 2000 & 10 & 5.5 \\
    & & AudioTime & 2000 & 10 & 5.5 \\    
    \cmidrule(lr){2-6}
    & \multirow{2}{*}{V2A} & VGGSound & 14,931 & 10 & 41.5 \\
    & & AVVP & 1,120 & 10 & 3.1 \\
    \cmidrule(lr){2-6}
    & TV2A & VGGSound & 14,931 & 10 & 41.5 \\
    \cmidrule(lr){2-6}
    & \multirow{2}{*}{T2M} & MusicCaps & 5,526 & 10 & 15.4 \\
    & & V2M & 3105 & 10 & 9.0 \\
    \cmidrule(lr){2-6}
    & V2M & V2M & 300 & 108 & 9.0 \\
    & TV2M & V2M & 300 & 108 & 9.0 \\
    \cmidrule(lr){2-6}    
    & \multirow{2}{*}{Audio Inpainting} & AudioCaps & 4,875 & 10 & 13.5 \\
    & & AVVP & 1,120 & 10 & 3.1 \\        
    \cmidrule(lr){2-6}
    & Music Completion & V2M & 300 & 108 & 9.0 \\    
    \bottomrule
    \end{tabular}}
    
    \label{tab:dataset_overview}
\end{table*}

%% file: table/dataset_2.tex
\begin{table}[h]
    \centering
    \small
    \renewcommand{\arraystretch}{1.2}
    \caption{Overview of our labeled captions, detailing the number of clips, average duration per clip, and total duration for each source dataset.}
    \resizebox{0.48\textwidth}{!}{
    \begin{tabular}{l c c c c}
    \toprule
    Source Dataset & Data Type & \# Clips & Dur./Clip (s) & Dur. (h) \\
    \midrule
    VGGSound & Audio & 191.8K & 10 & 532.81 \\
    AudioSet Strong & Audio & 67.3K & 10 & 187.14 \\
    AVVP Test Split & Audio & 1.1K & 10 & 3.11 \\
    V2M & Music & 7.9M & 10 & 21889.34 \\
    \bottomrule
    \end{tabular}}
    \label{tab:dataset}
\end{table}

%% file: table/avvp_v2a.tex
\begin{table}[!t]
    \centering
    \caption{
        \textbf{Performance evaluation on the AVVP dataset.} We report Video-to-Audio (V2A) and Text-and-Video-to-Audio (TV2A) results. For alignment (Align.), we use the ImageBind AV score for video inputs. Best per column is in \textbf{bold}, second best \underline{underlined}; cyan rows mark our methods.
    }
    \renewcommand{\arraystretch}{1.2}
    \setlength{\tabcolsep}{2.5pt}
    \footnotesize
    \adjustbox{max width=\columnwidth}{
    \begin{tabular}{ll ccccccc}
    \toprule
    Method & Task & KL $\downarrow$ & IS $\uparrow$ & FD $\downarrow$ & FAD $\downarrow$ & PC $\uparrow$ & PQ $\uparrow$ & Align.$\uparrow$ \\
    \midrule
    Seeing\&Hearing~\citep{xing2024seeing} & V2A  & 2.30 &  4.02 & 40.38 & 8.66 & 3.64 & 5.16 & \textbf{0.35} \\
    FoleyCrafter~\citep{zhang2024foleycrafter} & V2A & 2.13 & 6.46 & 28.68 & 3.77 & 3.25 & 5.87 & 0.28 \\
    Diff-Foley~\citep{luo2024diff} & V2A    & 3.14 &  5.97 & 76.96 & 10.95 & 2.55 & 5.71 & 0.16 \\
    FRIEREN~\citep{wang2024frieren} & V2A   & 2.73 &  4.71 & 66.46 & 6.49 & 3.08 & 5.88 & 0.17 \\
    MMAudio~\citep{cheng2025mmaudio} & V2A  & \textbf{1.22} & \underline{8.40} & \underline{13.51} & 3.25 & 3.55 & 5.89 & \underline{0.34} \\
    \cellcolor{cyan!7}\modelbase & \cellcolor{cyan!7}V2A & \cellcolor{cyan!7}4.55 & \cellcolor{cyan!7}\textbf{8.64} & \cellcolor{cyan!7}14.71 & \cellcolor{cyan!7}\underline{2.74} & \cellcolor{cyan!7}\underline{3.96} & \cellcolor{cyan!7}\textbf{6.15} & \cellcolor{cyan!7}0.30 \\
    \cellcolor{cyan!7}\model     & \cellcolor{cyan!7}V2A & \cellcolor{cyan!7}4.33 & \cellcolor{cyan!7}8.24 & \cellcolor{cyan!7}\textbf{13.05} & \cellcolor{cyan!7}\textbf{2.49} & \cellcolor{cyan!7}\textbf{4.15} & \cellcolor{cyan!7}\underline{6.08} & \cellcolor{cyan!7}0.31 \\
    \midrule
    FoleyCrafter~\citep{zhang2024foleycrafter} & TV2A & \underline{1.81} &  6.22 & 26.76 & 2.85 & 3.62 & 5.60 & 0.27 \\
    MMAudio~\citep{cheng2025mmaudio} & TV2A    & \textbf{1.74} & \textbf{9.52} & 14.18 & 2.74 & 3.64 & 5.81 & \textbf{0.34} \\
    \cellcolor{cyan!7}\modelbase & \cellcolor{cyan!7}TV2A & \cellcolor{cyan!7}3.39 & \cellcolor{cyan!7}\underline{9.38} & \cellcolor{cyan!7}\underline{12.85} & \cellcolor{cyan!7}\underline{2.38} & \cellcolor{cyan!7}\underline{3.90} & \cellcolor{cyan!7}\textbf{6.08} & \cellcolor{cyan!7}0.29 \\
    \cellcolor{cyan!7}\model     & \cellcolor{cyan!7}TV2A & \cellcolor{cyan!7}3.17 & \cellcolor{cyan!7}8.99 & \cellcolor{cyan!7}\textbf{12.22} & \cellcolor{cyan!7}\textbf{1.77} & \cellcolor{cyan!7}\textbf{4.08} & \cellcolor{cyan!7}\underline{6.03} & \cellcolor{cyan!7}\underline{0.31} \\
    \bottomrule
    \end{tabular}
    }
    \label{tab:avvp_v2a}
\end{table}

%% file: table/v2a_align.tex
\begin{table}[!t]
    \centering
    \caption{
        \textbf{Audio-visual alignment evaluation on the VGGSound V2A task.}
        In addition to standard quality metrics, we report two dedicated alignment metrics: AlignAcc (audio-visual semantic alignment accuracy, higher is better) and AVSync (audio-visual synchrony score, closer to 0 is better). Best per column is in \textbf{bold}, second best \underline{underlined}; cyan rows mark our methods.
    }
    \renewcommand{\arraystretch}{1.2}
    \setlength{\tabcolsep}{2.5pt}
    \footnotesize
    \adjustbox{max width=\columnwidth}{
    \begin{tabular}{l ccccccc cc}
    \toprule
    Method & KL $\downarrow$ & IS $\uparrow$ & FD $\downarrow$ & FAD $\downarrow$ & PC $\uparrow$ & PQ $\uparrow$ & Align.$\uparrow$ & AlignAcc $\uparrow$ & AVSync $\uparrow$ \\
    \midrule
    VATT~\citep{liu2024tell}                & \textbf{1.40} & 10.02 & 11.71 & 2.55 & 3.64 & 5.85 & 0.25 & 0.84 & -0.66 \\
    VAB~\citep{su2024vision}                & 2.30 &  8.15 & 20.21 & 3.05 & 3.52 & 5.93 & 0.24 & 0.82 & -0.64 \\
    MMAudio~\citep{cheng2025mmaudio}        & \underline{1.97} & \textbf{14.95} & \textbf{6.18} & 2.04 & 3.38 & 5.91 & \textbf{0.35} & 0.87 & \textbf{-0.56} \\
    \cellcolor{cyan!7}\modelbase            & \cellcolor{cyan!7}1.98 & \cellcolor{cyan!7}12.35 & \cellcolor{cyan!7}\underline{6.94} & \cellcolor{cyan!7}\underline{2.00} & \cellcolor{cyan!7}\underline{3.75} & \cellcolor{cyan!7}\textbf{6.31} & \cellcolor{cyan!7}0.28 & \cellcolor{cyan!7}\underline{0.89} & \cellcolor{cyan!7}-0.61 \\
    \cellcolor{cyan!7}\model                & \cellcolor{cyan!7}1.99 & \cellcolor{cyan!7}\underline{12.39} & \cellcolor{cyan!7}7.88 & \cellcolor{cyan!7}\textbf{1.34} & \cellcolor{cyan!7}\textbf{3.78} & \cellcolor{cyan!7}\underline{6.23} & \cellcolor{cyan!7}\underline{0.29} & \cellcolor{cyan!7}\textbf{0.90} & \cellcolor{cyan!7}\underline{-0.60} \\
    \bottomrule
    \end{tabular}
    }
    \label{tab:v2a_align}
\end{table}

%% file: table/audio_inpainting.tex
\begin{table}[h]
    \centering
    \footnotesize
    \caption{
        \textbf{Inpainting Performance Comparison.} This table shows the performance comparison for audio inpainting on the AudioCaps and AVVP datasets. The values before and after the slash represent the IS and FAD metrics, respectively. A, V, and T represent Audio, Video, and Text conditions. The baseline methods are all under audio and text conditions.
    }
    \resizebox{\columnwidth}{!}{    
    \begin{tabular}{l c c c}
    \toprule
    \multirow{2.5}{*}{Method} & \multirow{2.5}{*}{Input} & \multicolumn{2}{c}{Dataset} \\
    \cmidrule(lr){3-4}
     &  & AudioCaps & AVVP \\
    \midrule
    Unprocessed & - & 6.51/11.34 & 4.94/6.70 \\
    AudioLDM-L-Full\citep{liu2024audioldm} & A+T & 8.06/2.64 & 5.11/3.30 \\
    AudioLDM-2-Full-Large\citep{liu2024audioldm} & A+T & 4.24/10.17 & 3.99/11.58 \\    
    \midrule
    \cellcolor{cyan!7}\modelbase & \cellcolor{cyan!7}A     & \cellcolor{cyan!7}4.47/7.10 & \cellcolor{cyan!7}3.91/5.40 \\
    \cellcolor{cyan!7}\modelbase & \cellcolor{cyan!7}A+T   & \cellcolor{cyan!7}8.59/2.66 & \cellcolor{cyan!7}5.75/2.14 \\
    \cellcolor{cyan!7}\modelbase & \cellcolor{cyan!7}A+V   & \cellcolor{cyan!7}N/A & \cellcolor{cyan!7}5.20/2.42 \\
    \cellcolor{cyan!7}\modelbase & \cellcolor{cyan!7}A+T+V & \cellcolor{cyan!7}N/A & \cellcolor{cyan!7}5.79/2.39 \\
    \cmidrule(lr){1-4}
    \cellcolor{cyan!7}\model & \cellcolor{cyan!7}A     & \cellcolor{cyan!7}4.92/6.11 & \cellcolor{cyan!7}3.98/5.61 \\
    \cellcolor{cyan!7}\model & \cellcolor{cyan!7}A+T   & \cellcolor{cyan!7}8.27/2.82 & \cellcolor{cyan!7}5.18/2.24 \\
    \cellcolor{cyan!7}\model & \cellcolor{cyan!7}A+V   & \cellcolor{cyan!7}N/A & \cellcolor{cyan!7}4.85/2.94 \\
    \cellcolor{cyan!7}\model & \cellcolor{cyan!7}A+T+V & \cellcolor{cyan!7}N/A & \cellcolor{cyan!7}5.99/2.52 \\
    \bottomrule
    \end{tabular}
    } 

    \label{tab:inpainting_results}
\end{table}

%% file: table/music_comp.tex
\begin{table}[h]
    \centering
    \small
    \caption{\textbf{Performance for our method under different conditions in the music completion task.} M, T, and V represent Music, Text, and Video, respectively.}
    \resizebox{0.4\textwidth}{!}{        
    \begin{tabular}{c c c c c}
        \toprule
        Input & KL $\downarrow$ & IS $\uparrow$ & FD $\downarrow$ & FAD $\downarrow$ \\
        \midrule
        M & 0.96 & 1.21 & 52.77 & 5.76 \\
        T+M & \underline{0.51} & \underline{1.49} & \underline{21.42} & \underline{2.14} \\
        V+M & 0.70 & 1.37 & 24.28 & 2.29 \\
        T+V+M & \textbf{0.46} & \textbf{1.52} & \textbf{18.69} & \textbf{1.67} \\
        \bottomrule
    \end{tabular}
    }       
    
    \label{tab:music_comp}
\end{table}

%% file: table/I2A.tex
\begin{table}[h]
    \centering
    \small
    \caption{
        \textbf{Comparison of Methods for the Image2Audio Task.}
    }
    \resizebox{\columnwidth}{!}{
    \begin{tabular}{c c c c c c}
    \toprule
    Method & KL $\downarrow$ & IS $\uparrow$ & FD $\downarrow$ & FAD $\downarrow$ & Align. $\uparrow$ \\
    \midrule
    Caption2Audio & 2.76 & 7.48 & 32.97 & 5.54 & 0.21 \\
    Im2Wav\citep{sheffer2023hear} & \textbf{2.61} & 7.06 & 19.63 & 7.58 & \textbf{0.41} \\  
    Seeing\&Hearing\citep{xing2024seeing} & \underline{2.69} & 6.15 & 20.96 & 6.87 & \underline{0.29} \\
    \cellcolor{cyan!7}\modelbase & \cellcolor{cyan!7}2.81 & \cellcolor{cyan!7}\textbf{11.47} & \cellcolor{cyan!7}\underline{17.03} & \cellcolor{cyan!7}\underline{2.68} & \cellcolor{cyan!7}0.22 \\
    \cellcolor{cyan!7}\model & \cellcolor{cyan!7}2.85 & \cellcolor{cyan!7}\underline{11.22} & \cellcolor{cyan!7}\textbf{16.53} & \cellcolor{cyan!7}\textbf{2.65} & \cellcolor{cyan!7}0.23 \\
    \bottomrule
    \end{tabular}
    }
    
    \label{tab:image2audio_comparison}
\end{table}